\documentclass[amstex,12pt]{article}
\usepackage{amssymb}
\usepackage{amsmath}
\usepackage{IEEEtrantools}
\usepackage{pdflscape}
\usepackage{afterpage}
\usepackage{capt-of}

\usepackage{graphicx}
\usepackage{epstopdf}
\usepackage{subfig}

\usepackage{changepage}

\numberwithin{equation}{section}

\newcommand{\be}{\begin{equation}}
\newcommand{\ee}{\end{equation}}
\newcommand{\bea}{\begin{eqnarray}}
\newcommand{\eea}{\end{eqnarray}}
\newcommand{\non}{\nonumber}
\newcommand{\id}{\mathbb{I}}

\newcommand{\tr}{\mathop{\rm tr}\nolimits}
\newcommand{\diag}{\mathop{\rm diag}\nolimits}
\newcommand{\uh}{\frac{u}{2}}

\newcommand{\mk}{\mathsf{k}}

\usepackage[usenames,dvipsnames]{color}

\begin{document}

\begin{titlepage}
\strut\hfill UMTG--291
\vspace{.5in}
\begin{center}

\LARGE The integrable quantum group invariant\\
$A_{2n-1}^{(2)}$ and $D_{n+1}^{(2)}$ open spin chains\\
\vspace{1in}
\large 
Rafael I. Nepomechie \footnote{Physics Department,
P.O. Box 248046, University of Miami, Coral Gables, FL 33124 USA, nepomechie@miami.edu}
, Rodrigo A. Pimenta \footnote{Instituto de F\'{i}sica de S\~{a}o Carlos, Universidade de S\~{a}o Paulo, Caixa
Postal 369, 13560-590, S\~{a}o Carlos, SP, Brazil, pimenta@ifsc.usp.br}, 
and Ana L. Retore \footnote{Instituto de F\'{i}sica Te\'{o}rica-UNESP, Rua 
Dr. Bento Teobaldo Ferraz 271, Bloco II 01140-070, S\~{a}o Paulo, Brazil, retore@ift.unesp.br}\\[0.8in]
\end{center}

\vspace{.5in}

\begin{abstract}
A family of $A_{2n}^{(2)}$ integrable open spin chains with
$U_{q}(C_{n})$ symmetry was recently identified in arXiv:1702.01482.
We identify here in a similar way a family of $A_{2n-1}^{(2)}$
integrable open spin chains with $U_{q}(D_{n})$ symmetry, and two
families of $D_{n+1}^{(2)}$ integrable open spin chains with
$U_{q}(B_{n})$ symmetry.  We discuss the consequences of these
symmetries for the degeneracies and multiplicities of the spectrum.
We propose Bethe ansatz solutions for two of these models, whose
completeness we check numerically for small values of $n$ and chain
length $N$.  We find formulas for the Dynkin labels in terms of the
numbers of Bethe roots of each type, which are useful for determining
the corresponding degeneracies. In an appendix, we 
briefly consider $D_{n+1}^{(2)}$ chains with other integrable 
boundary conditions, which do not have quantum group symmetry. 
\end{abstract}

\end{titlepage}

\setcounter{footnote}{0}

\section{Introduction}\label{sec:intro}

Quantum spin chains are quantum many-body systems that have
applications in diverse fields, ranging from statistical mechanics 
\cite{Baxter1982, McCoy2010}, condensed-matter theory \cite{Giamarchi2004, Mikeska:2004} 
and quantum information theory \cite{Marchukov2016} to quantum
field theory and string theory \cite{Beisert:2010jr}.  The simplest anisotropic spin chains
are arguably those that are integrable and have quantum group \cite{Chari:1994pz}
symmetry.  Indeed, integrability allows access to the spectrum, and
quantum group symmetry can account for the degeneracies and
multiplicities.  The prototypical example is the
$U_{q}(A_{1})$-invariant open spin-1/2 chain \cite{Pasquier:1989kd},
whose integrability follows from \cite{Alcaraz:1987uk,
Sklyanin:1988yz}.

Generalizations of this example can be constructed systematically.
Integrable bulk interactions are encoded in solutions of the
Yang-Baxter equation, which in this context are called R-matrices.
Infinite families of anisotropic R-matrices associated with
corresponding affine Lie algebras were found in \cite{Bazhanov:1984gu,
Bazhanov:1986mu, Jimbo:1985ua, Kuniba:1991yd}.  Similarly, integrable
boundary conditions are encoded in solutions of the boundary
Yang-Baxter equation (BYBE) \cite{Sklyanin:1988yz, Cherednik:1985vs,
Ghoshal:1993tm}, which in this context are called $K^{\mp}$-matrices.
Quantum group symmetry can be realized by choosing these 
$K^{\mp}$-matrices appropriately.

Several infinite families of integrable open spin chains that are
quantum group invariant were identified in \cite{Mezincescu:1990ui}.
Bethe ansatz solutions of these models were found in
\cite{Mezincescu:1991ag, Artz:1994qy, Artz:1995bm, Li:2005pp,
Li:2006mv}.  The integrable quantum-group-invariant spin chains
identified in \cite{Mezincescu:1990ui} are all constructed using the
R-matrices in \cite{Jimbo:1985ua, Kuniba:1991yd} together with the {\em
simplest} $K^{-}$-matrix, namely, the identity matrix.  For example,
the integrable spin chain constructed with the $A_{2n}^{(2)}$ R-matrix
and $K^{-}=\id$ has $U_{q}(B_{n})$ symmetry.  The appearance of
$B_{n}$ can be understood from the fact (see e.g. \cite{Kac:1983})
that it is the subalgebra of $A_{2n}$ that remains invariant under the
order-2 diagram automorphism.

\begin{figure}[b]
\centering
\subfloat[]{\includegraphics[width=5.5cm]{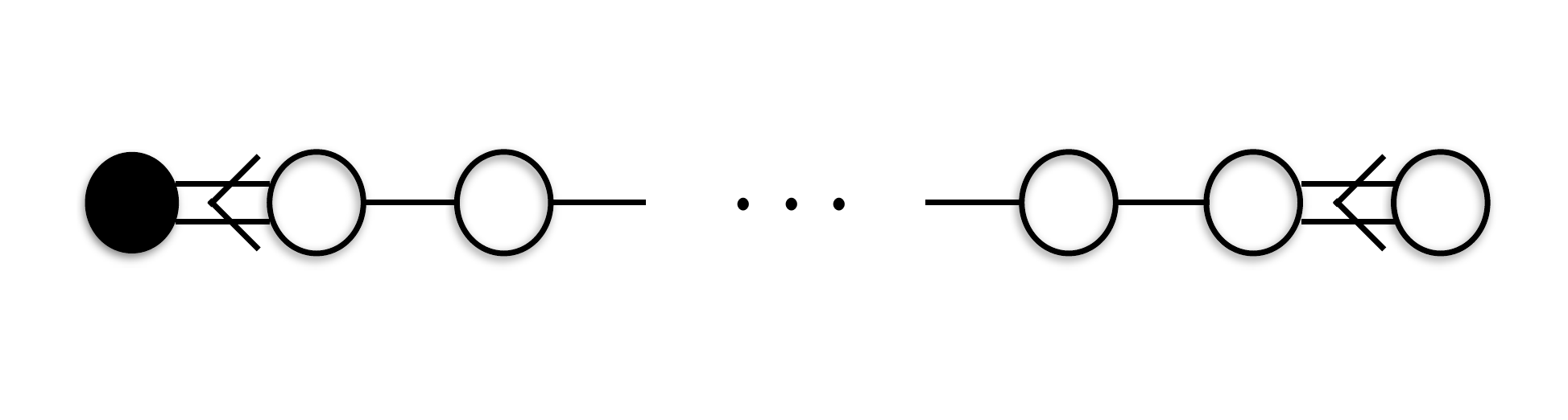}}
\subfloat[]{\includegraphics[width=5.5cm]{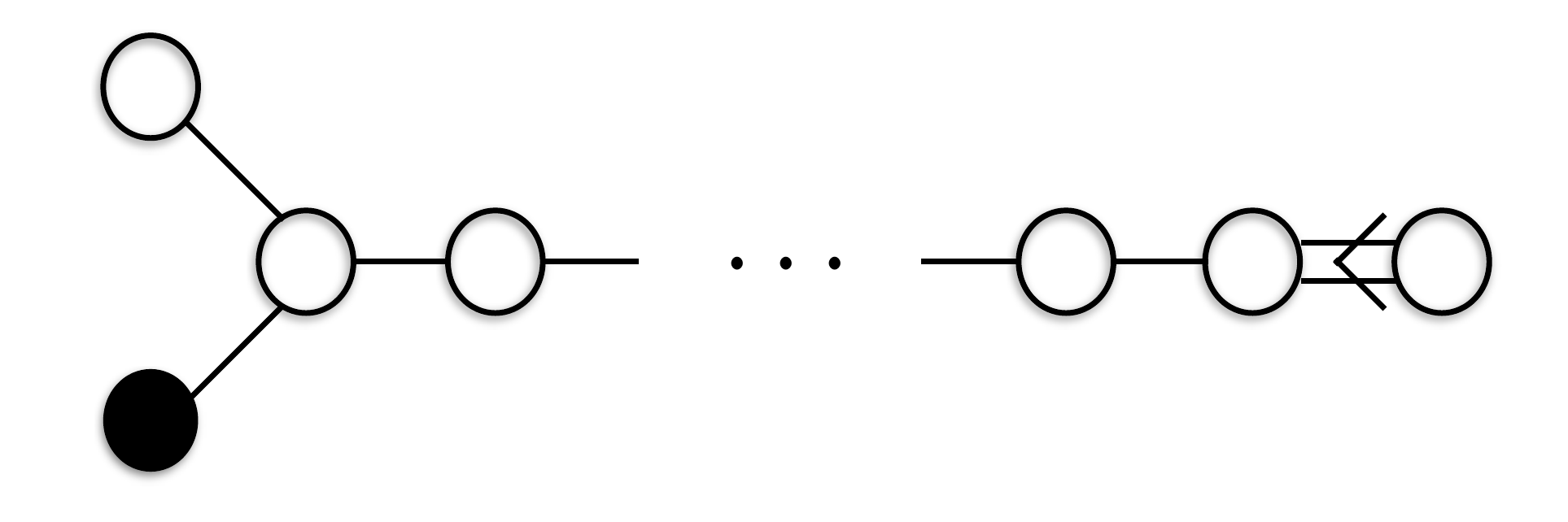}}
\subfloat[]{\includegraphics[width=5.5cm]{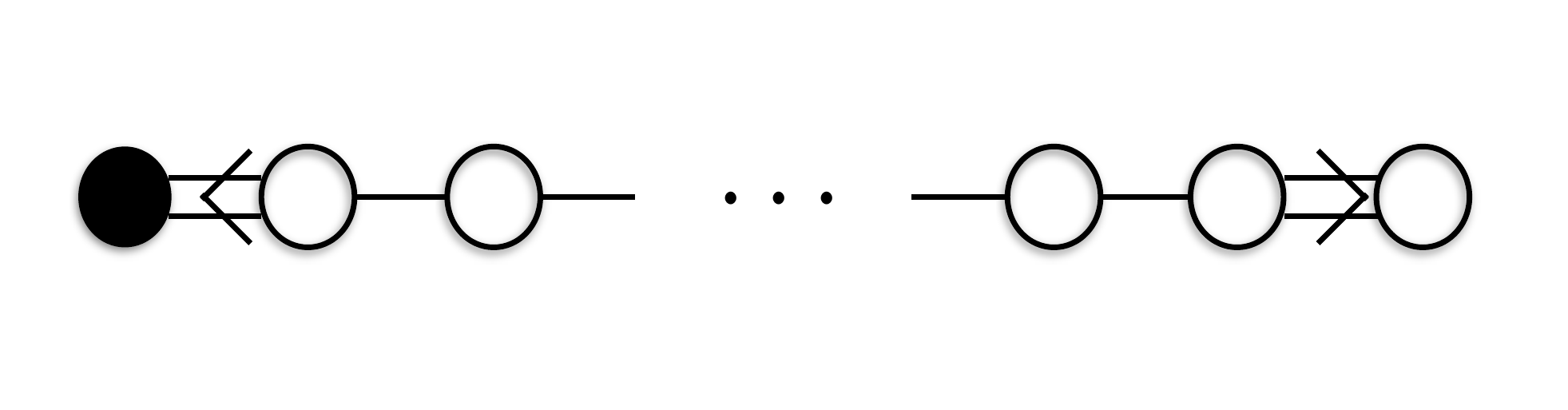}}
\caption{Extended Dynkin diagrams for  
(a) $A^{(2)}_{2n}$ (b) $A^{(2)}_{2n-1}$ (c) $D^{(2)}_{n+1}$}
\label{fig:extendedDynkin}
\end{figure}

\begin{figure}
\centering
\subfloat[]{\includegraphics[width=5.5cm]{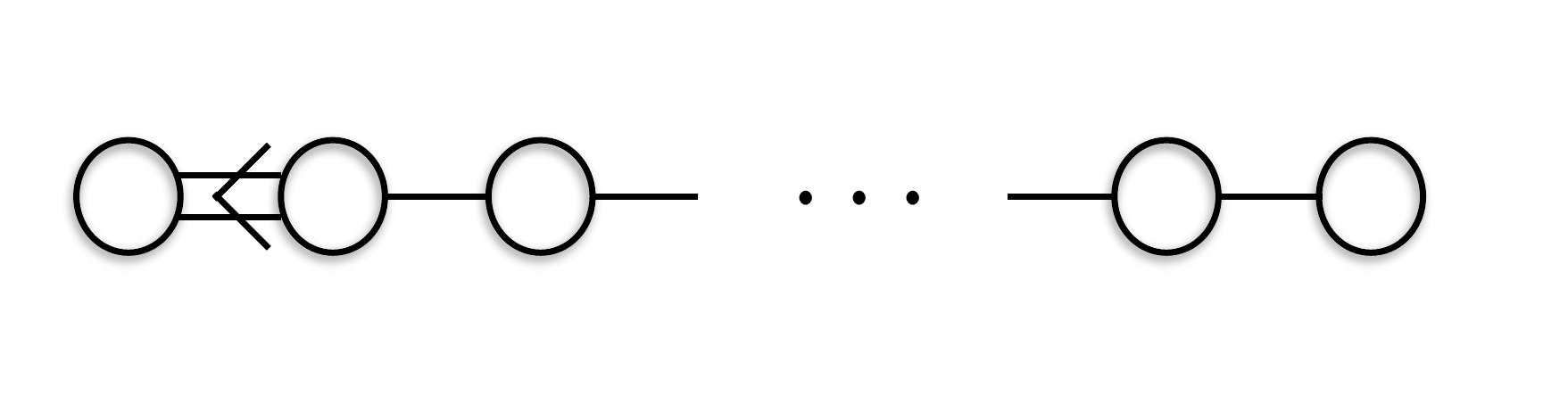}}
\subfloat[]{\includegraphics[width=5.5cm]{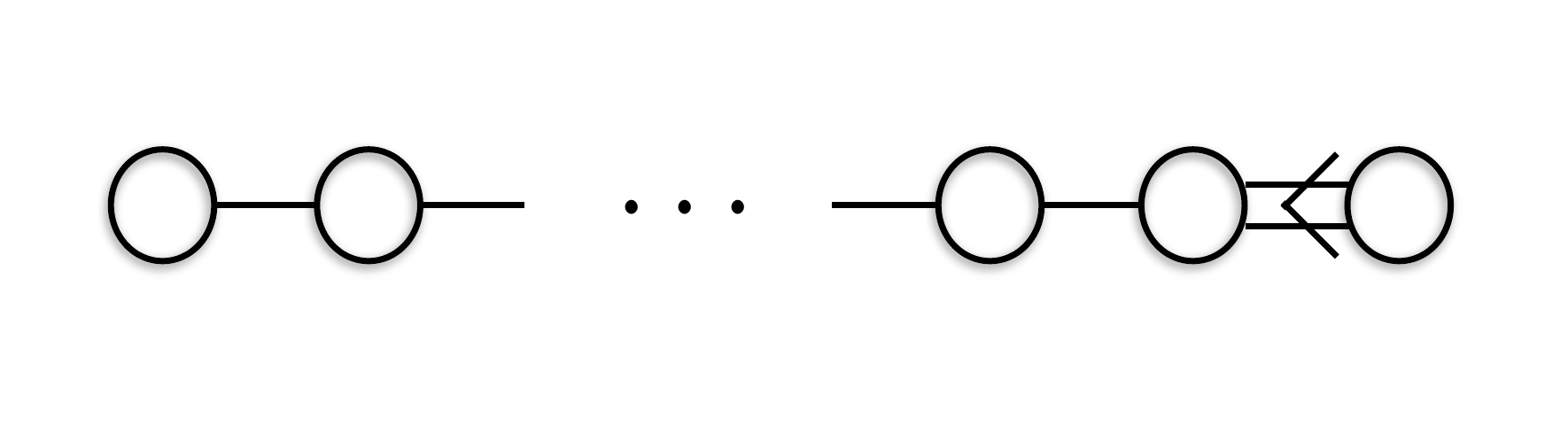}}
\subfloat[]{\includegraphics[width=5.5cm]{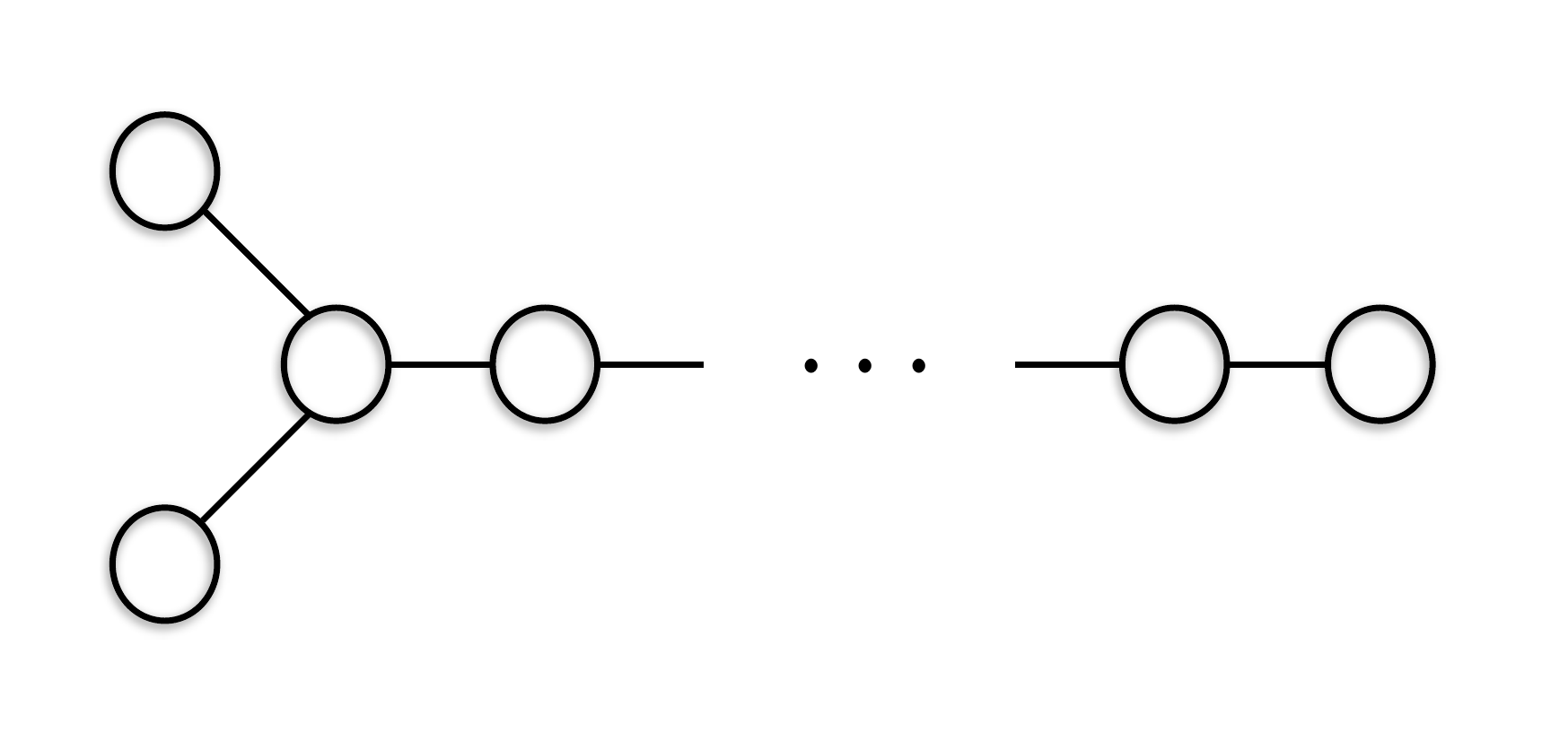}}
\caption{Dynkin diagrams for  
(a) $B_{n}$ (b) $C_{n}$ (c) $D_{n}$}
\label{fig:Dynkin}
\end{figure}

It was recently observed that the integrable spin chain constructed
with the $A_{2n}^{(2)}$ R-matrix and with a {\em different} choice of $K^{-}$
\cite{Mezincescu:1990ui, Batchelor:1996np, LimaSantos:2002ui} has
instead $U_{q}(C_{n})$ symmetry \cite{Ahmed:2017mqq}.  (The case
$n=1$, corresponding to the Izergin-Korepin R-matrix
\cite{Izergin:1980pe}, was analyzed in \cite{Nepomechie:1999jz}.)
The appearance of $C_{n}$ can be inferred from the extended Dynkin
diagram for $A_{2n}^{(2)}$ shown in Fig. \ref{fig:extendedDynkin}: 
removing the leftmost node yields the Dynkin diagram for the 
subalgebra $C_{n}$ shown in Fig \ref{fig:Dynkin}.  (Removing the
rightmost node yields the Dynkin diagram for the subalgebra $B_{n}$; and indeed the
spin chain with $K^{-}=\id$ has $U_{q}(B_{n})$ symmetry, as already 
noted above.)

This observation in \cite{Ahmed:2017mqq} opens the door to (at least)
doubling the list of integrable quantum-group-invariant spin chains
identified in \cite{Mezincescu:1990ui}: starting with an R-matrix
corresponding to a given affine Lie algebra, it should be possible to
construct a different integrable spin chain that is invariant under the
quantum group dictated by the corresponding extended Dynkin diagram 
(as illustrated above) by finding an appropriate $K^{-} \ne \id$.

We carry out here the above program for the R-matrices corresponding
to the two remaining infinite families of twisted affine Lie algebras,
namely $A_{2n-1}^{(2)}$ and $D_{n+1}^{(2)}$.  For $A_{2n-1}^{(2)}$ with $n>1$, we
expect to find a new integrable spin chain with $U_{q}(D_{n})$
symmetry, since removing the rightmost node of the extended Dynkin
diagram (see again Fig.  \ref{fig:extendedDynkin}) yields the Dynkin
diagram for the subalgebra $D_{n}$.  (Removing one of the two leftmost
nodes yields the Dynkin diagram for the subalgebra $C_{n}$, and indeed
the spin chain with $K^{-}=\id$ has $U_{q}(C_{n})$ symmetry \cite{Mezincescu:1990ui}.)

Integrable quantum-group-invariant spin chains with the
$D_{n+1}^{(2)}$ R-matrix were not constructed in
\cite{Mezincescu:1990ui}, since the corresponding BYBE does {\em not}
have the solution $K^{-}=\id$.  However, several solutions of this BYBE
were found in \cite{Martins:2000xie}.  We show here that two of these
solutions can be used to construct integrable spin chains with
$U_{q}(B_{n})$ symmetry, as expected from the extended Dynkin diagram
for $D_{n+1}^{(2)}$ (see again Fig.  \ref{fig:extendedDynkin}), since
removing either the leftmost or rightmost node yields the Dynkin
diagram for the subalgebra $B_{n}$.

The outline of this paper is as follows.  In Sec.  \ref{sec:Rmatgen}
we recall basic properties of the $A_{2n-1}^{(2)}$ and $D_{n+1}^{(2)}$
R-matrices, whose explicit expressions are given in
Appendix \ref{sec:Rmatexplicit}.  In
Sec.  \ref{sec:Kmat} we introduce a pair of K-matrices for each
R-matrix, which in a subsequent section will be shown to lead to
quantum group symmetry.  We briefly review in Sec.  \ref{sec:transfer}
how the R-matrices and K-matrices can be used to construct the
transfer matrix, and consequently the Hamiltonian, of an integrable
open spin chain.  We then use an identity (\ref{identity2}) to show
that -- for each of our four choices of K-matrices -- the
corresponding Hamiltonians can be expressed essentially as sums of
two-body terms, see (\ref{HamiltonianA2oddI}),
(\ref{HamiltonianA2oddII}), (\ref{HamiltonianD2I}) and
(\ref{HamiltonianD2II}).  We use this fact in Sec.  \ref{sec:QG} to
show that each of these four Hamiltonians has quantum group symmetry.
We also discuss the consequences of these symmetries for the degeneracies
and multiplicities of the spectrum. In Secs. \ref{sec:BAA2odd} and 
\ref{sec:BAD2} we present Bethe ansatz solutions for all but one of 
these models, and numerically check their completeness.\footnote{By 
``completeness'' we mean here that the Bethe ansatz produces all of the 
eigenvalues of the transfer matrix.} We also find formulas 
for the Dynkin labels in terms of the numbers of Bethe roots of each 
type, which are useful for determining the corresponding degeneracies.
In Sec. \ref{sec:conclusion} we briefly summarize our main results and 
note some remaining open problems. In Appendix \ref{sec:other}, we 
briefly consider $D_{n+1}^{(2)}$ chains with other integrable 
boundary conditions, which do not have quantum group symmetry.

\section{R-matrices: generalities}\label{sec:Rmatgen}

The R-matrices for $A_{2n-1}^{(2)}$ and $D_{n+1}^{(2)}$ are given 
explicitly in 
Appendix \ref{sec:Rmatexplicit}. These R-matrices map ${\cal V} \otimes {\cal V}$ to itself, where 
${\cal V}$ is a $d$-dimensional vector space, where
\be
d = \left\{ \begin{array}{cc}
2n & \mbox{ for  } A_{2n-1}^{(2)} \\
2n+2 & \mbox{ for  } D_{n+1}^{(2)} 
\end{array} \right. \,,
\ee  
and satisfy the Yang-Baxter equation (YBE) on ${\cal V} \otimes {\cal V}\otimes {\cal V}$
\be
R_{12}(u - v)\,  R_{13}(u)\, R_{23}(v) = R_{23}(v)\,  R_{13}(u)\, R_{12}(u - v)
\,.  \label{YBE}
\ee
As usual, $R_{12} = R \otimes \id\,, R_{23} = \id \otimes R\,, R_{13} = 
{\cal P}_{23}  R_{12} {\cal P}_{23}$, where $\id$ is the identity 
matrix on ${\cal V}$, and ${\cal P}$ is the permutation matrix on ${\cal V} \otimes {\cal V}$
\be
{\cal P}=\sum_{\alpha, \beta = 1}^d e_{\alpha \beta}\otimes e_{\beta 
\alpha} \,,
\ee
and $e_{\alpha \beta}$ are the $d \times d$ elementary 
matrices with elements $(e_{\alpha \beta})_{ij} = 
\delta_{\alpha, i} \delta_{\beta, j}$. In addition, these R-matrices 
have the following properties: $PT$ symmetry
\be
R_{21}(u) \equiv {\cal P}_{12}\, R_{12}(u)\, {\cal P}_{12} 
= R_{12}^{t_1 t_2}(u) \,,
\label{PT}
\ee
unitarity
\be
R_{12}(u)\ R_{21}(-u) = \zeta(u)\, \id\otimes\id  \,,
\label{unitarity}
\ee
where $\zeta(u)$ is given by
\be
\zeta(u) =\xi(u)\, \xi(-u)\,, \qquad 
\xi(u)=\left\{ \begin{array}{cc}
-2 \sinh(\frac{u}{2} +2\eta) \cosh(\frac{u}{2} +2 n \eta)&  \mbox{ for  } A_{2n-1}^{(2)} \\
4\sinh(u +2\eta) \sinh(u +2 n \eta)  & \mbox{ for  } D_{n+1}^{(2)} 
\end{array} \right. \,,
\label{xi}	    
\ee
regularity
\be
R(0) = \xi(0)\, {\cal P}\,,
\label{Rregular}
\ee
and crossing symmetry
\be
R_{12}(u)=V_1\, R_{12}^{t_2}(-u-\rho)\, V_1
= V_2^{t_2}\, R_{12}^{t_1}(-u-\rho)\, V_2^{t_2} \,,
\label{crossing}
\ee
where the crossing parameter $\rho$ is given by
\be
\rho= \left\{ \begin{array}{cc}
-4 n\eta - i \pi &  \mbox{ for  } A_{2n-1}^{(2)} \\
-2 n\eta & \mbox{ for  } D_{n+1}^{(2)} 
\end{array} \right. \,.
\ee
The crossing matrix $V$ is given by
\be
V= \sum_{k=1}^{d} v_{k}\, e_{k,d+1-k} \,, 
\ee
where
\begin{align} 
v_{k} &=
\left\{ \begin{array}{cl}
i e^{-2(n+1-k)\eta} & k = 1, \ldots, n \\
-i e^{-2(n-k)\eta} & k = n+1, \ldots , 2n
\end{array} \right. \mbox{ for  } A_{2n-1}^{(2)} \\
v_{k} &= \left\{ \begin{array}{cl}
e^{-(2n+1-2k)\eta} & k = 1, \ldots, n \\
1 & k = n+1, n+2 \\
e^{-(2n+5-2k)\eta} & k = n+3, \ldots , 2n+2
\end{array} \right. \mbox{ for  } D_{n+1}^{(2)}  \,,
\end{align}
and satisfies $V^2 = \id$. The corresponding matrix $M$ is defined by
\be
M = \epsilon V^{t}\, V\,, \qquad \epsilon =\left\{ \begin{array}{cc}
-1 &  \mbox{ for  } A_{2n-1}^{(2)} \\
+1 & \mbox{ for  } D_{n+1}^{(2)} 
\end{array} \right. \,,
\label{Mdef}
\ee
and is given by a diagonal matrix
\be
M  =\left\{ \begin{array}{ll}
\diag(e^{4(n+\frac{1}{2}-\bar\alpha)\eta})\,, \qquad \alpha = 1, 
2, \ldots \,, 2n &  \mbox{ for  } A_{2n-1}^{(2)} \\
\diag(e^{4(n+\frac{3}{2}-\bar\alpha)\eta})\,, \qquad \alpha = 1, 
2, \ldots \,, 2n+2 & \mbox{ for  } D_{n+1}^{(2)} 
\end{array} \right. 
\label{Mmat}
\ee
where $\bar\alpha$ is defined by (\ref{alphabarA2odd}) for $A_{2n-1}^{(2)}$ and by  (\ref{alphabarD2}) for $D_{n+1}^{(2)}$.
 
\section{K-matrices}\label{sec:Kmat}

The matrix $K^{-}(u)$, which maps ${\cal V}$ to itself, is a solution of the 
so-called reflection equation or boundary Yang-Baxter equation (BYBE) on ${\cal V} \otimes {\cal V}$ 
\cite{Sklyanin:1988yz, Cherednik:1985vs, Ghoshal:1993tm}
\be
R_{12}(u - v)\, K^-_1(u)\ R_{21} (u + v)\, K^-_2(v)
= K^-_2(v)\, R_{12}(u + v)\, K^-_1(u)\, R_{21}(u - v)  \,.
\label{BYBEm}
\ee
We assume that it has the regularity property
\be
K^{-}(0) = \kappa\, \id \,.
\label{Kregular}
\ee

For the corresponding $K^{+}$-matrix, we shall always take
\be
K^{+}(u) = K^{-\, t}(-u-\rho)\, M\,,
\label{isomorphism}
\ee
where $M$ is defined by (\ref{Mdef}).
For any solution $K^{-}(u)$ of the BYBE (\ref{BYBEm}), 
the ``isomorphism'' (\ref{isomorphism}) gives a solution of the 
corresponding BYBE for $K^{+}(u)$ \cite{Sklyanin:1988yz, Mezincescu:1990uf}
\begin{eqnarray}
\lefteqn{R_{12}(-u + v)\, K_1^{+\, t_1}(u)\, M^{-1}_1\, R_{21} (-u -v 
-2\rho)\, M_1\, K_2^{+\, t_2}(v)} \non \\
& & \quad = K^{+\, t_2}_2(v)\, M_1\, R_{12}(-u - v- 2\rho)\, M^{-1}_1\,
K^{+\, t_1}_1(u)\, R_{21}(-u +v)  \,.
\label{BYBEp}
\end{eqnarray} 

\subsection{$A_{2n-1}^{(2)}$ K-matrices}

For the BYBE (\ref{BYBEm}) with Kuniba's 
$A_{2n-1}^{(2)}$ R-matrix (\ref{RA2odd})-(\ref{alphabeta}), only two 
solutions have (to our knowledge) been reported \cite{Artz:1995bm, 
Batchelor:1996np}, both of which are diagonal.\footnote{In contrast, for the BYBE with Jimbo's 
$A_{2n-1}^{(2)}$ R-matrix \cite{Jimbo:1985ua}, more solutions are 
known \cite{LimaSantos:2003hx, Malara:2004bi}.} We have 
searched for additional diagonal solutions with $n \ge 2$, and we have 
found only one more:
\be
K^{-}(u) = \diag( \frac{e^{-u} + \beta}{e^{u} + \beta}, 1, 1, 
\ldots, 1, 1,  \frac{e^{u + 4 (n-1) \eta} - \beta}{e^{-u + 4 (n-1) 
\eta} - \beta}) \,,
\ee
where $\beta$ is an arbitrary parameter. This solution reduces to the identity matrix 
$\id_{2n \times 2n}$ in the limit $\beta \rightarrow \infty$. We have not looked for non-diagonal solutions, which would 
require significantly more effort.

For $A_{2n-1}^{(2)}$, we shall henceforth consider the following two $2n \times 2n$
diagonal $K^{-}$-matrices, to which we refer by I and II:
\begin{align}
    A_{2n-1}^{(2)} - \text{I}:  & \qquad K^{-}(u) = \id_{2n \times 2n}\,, 
\label{KA2oddsetI} \\
   A_{2n-1}^{(2)} - \text{II}: & \qquad K^{-}(u) = \left( \begin{array}{cc}
e^{-u}\, \id_{n \times n} & \\
& e^{u}\, \id_{n \times n}
\end{array} \right) \,.
\label{KA2oddsetII}
\end{align}
These two solutions of the $A_{2n-1}^{(2)}$ BYBE (\ref{BYBEm}) were 
found in \cite{Mezincescu:1990ui, Artz:1995bm} 
and \cite{Batchelor:1996np}, respectively. Both solutions evidently 
satisfy the regularity property (\ref{Kregular}) with 
\be
\kappa = 1 \,.
\label{kappaA2odd}
\ee

\subsection{$D_{n+1}^{(2)}$ K-matrices}

Several solutions of the BYBE (\ref{BYBEm}) with the 
$D_{n+1}^{(2)}$ R-matrix (\ref{RD2}) are known \cite{Martins:2000xie, 
Malara:2004bi}. A diagonal solution is given by the following $(2n+2) \times 
(2n+2)$ matrix \cite{Martins:2000xie}
\be
K^{-}(u) = \left( \begin{array}{cccc}
\id_{n \times n} & & & \\
& \mk_{1}(u) &  & \\
& & \mk_{2}(u) & \\
& & & e^{2u}\, \id_{n \times n}
\end{array} \right) \,,
\label{KD2diag}
\ee
where
\be
\mk_{1}(u) = \frac{e^{u}- i e^{- n \eta}}{e^{-u}- i e^{- n \eta}}\,, 
\qquad 
\mk_{2}(u) = \frac{e^{u}+ i e^{- n \eta}}{e^{-u}+ i e^{- n \eta}}\,.
\label{KD2diag2}
\ee
We consider this solution briefly in Appendix \ref{sec:diag}. 

Two block-diagonal solutions of the $D_{n+1}^{(2)}$ BYBE are also
known \footnote{See Eqs.
(57)-(60) and (61)-(64) in 
\cite{Martins:2000xie}. The notations in 
\cite{Martins:2000xie} are related to 
ours by $\lambda=u$, $q=e^{2\eta}$, and $n_{\textrm{\cite{Martins:2000xie}}} = 
n+1$. Moreover, in the second K-matrix, we make the replacement 
$\sqrt{\xi_{-}} \mapsto \xi_{-}$.
It was conjectured in \cite{Martins:2000xie} that these 
K-matrices with $\xi_{-}=0$ lead to quantum group symmetry.}
: one of these solutions is given by the 
$(2n+2) \times (2n+2)$ matrix
\be 
K^{-}(u) = 
\left( \begin{array}{cccc}
\mk_{0}(u)\, \id_{n \times n} & & & \\
& \mk_{1}(u) & \mk_{2}(u) & \\
& \mk_{3}(u) & \mk_{4}(u) & \\
& & & \mk_{5}(u)\, \id_{n \times n}
\end{array} \right) \,,   
\label{KD2blockparamI1}
\ee 
where
\begin{align}
\mk_{0}(u) &= (e^{2u}+e^{2 n \eta})(\xi_{-}^{2} e^{u +  2n \eta} - e^{-u})\,, \non\\
\mk_{1}(u) &= \frac{1}{2}(e^{2u}+1)\left[2 \xi_{-} e^{2 n \eta}(e^{2u}-1) - 
e^{u}(1-\xi_{-}^{2} e^{2 n \eta})(1 + e^{2 n \eta}) \right] \,, \non\\
\mk_{2}(u) &= \mk_{3}(u) = \frac{1}{2}e^{u} (e^{2u}-1) (1+\xi_{-}^{2} e^{2 n 
\eta})(1 - e^{2 n \eta}) \,, \non\\
\mk_{4}(u) &= \frac{1}{2}(e^{2u}+1)\left[-2 \xi_{-} e^{2 n \eta}(e^{2u}-1) - 
e^{u}(1-\xi_{-}^{2} e^{2 n \eta})(1 + e^{2 n \eta}) \right] \,, \non\\
\mk_{5}(u) &= (e^{2u}+e^{2 n \eta})(\xi_{-}^{2} e^{u +  2n \eta} - e^{3u})\,,
\label{KD2blockparamI2}
\end{align}
and $\xi_{-}$ is an arbitrary boundary parameter. We consider this 
solution briefly in Appendix \ref{sec:blockdiag}. Another 
block-diagonal solution has the same matrix structure 
(\ref{KD2blockparamI1}), but the matrix elements are instead given by
\begin{align}
\mk_{0}(u) &= (e^{2u}-e^{2 n \eta})(\xi_{-}^{2} e^{u}-e^{-u})  \,, \non\\
\mk_{1}(u) &= \mk_{4}(u) = \frac{1}{2}(e^{2u}+1) e^{u} (1 - e^{2 n \eta})(\xi_{-}^{2} - 1) \,, \non\\
\mk_{2}(u) &= \frac{1}{2}(e^{2u}-1)\left[ 2 e^{n \eta} (e^{2u}+1) 
\xi_{-} + e^{u} (1 + e^{2 n \eta})(1 + \xi_{-}^{2}) \right] \,, \non\\
\mk_{3}(u) &= \frac{1}{2}(e^{2u}-1)\left[ -2 e^{n \eta} (e^{2u}+1) 
\xi_{-} + e^{u} (1 + e^{2 n \eta})(1 + \xi_{-}^{2}) \right] \,, \non\\
\mk_{5}(u) &= (e^{2u}-e^{2 n \eta})(\xi_{-}^{2} - e^{2u}) e^{u}\,,
\label{KD2blockparamII}
\end{align}
where $\xi_{-}$ is again an arbitrary boundary parameter.

For $\xi_{-} = 0$, these block-diagonal solutions have the 
matrix structure
\be 
K^{-}(u) = 
-2 e^{2u+ n \eta}\left( \begin{array}{cccc}
k_{-}(u)\, \id_{n \times n} & & & \\
& k_{1}(u) & k_{2}(u) & \\
& k_{2}(u) & k_{1}(u) & \\
& & & k_{+}(u)\, \id_{n \times n}
\end{array} \right) \,,   
\label{KD2}
\ee 
and the matrix elements are given by
\begin{align}
D_{n+1}^{(2)} - \text{I}:  & \qquad 
k_{\mp}(u) = e^{\mp 2u}\cosh(u- n \eta)\,,\quad 
k_{1}(u) = \cosh(u)  \cosh(n \eta)\,,   \non  \\
& \qquad\qquad k_{2}(u) = \sinh(u)  \sinh(n \eta) \,, \label{QGKMI} \\
D_{n+1}^{(2)} - \text{II}:  & \qquad
k_{\mp}(u) = e^{\mp 2u}\sinh(u- n \eta)\,,\quad 
k_{1}(u) = -\cosh(u)  \sinh(n \eta)\,,  \non  \\
& \qquad \qquad k_{2}(u) = -\sinh(u)  \cosh(n \eta) \,.
\label{QGKMII} 
\end{align}
corresponding to (\ref{KD2blockparamI2}) and  
(\ref{KD2blockparamII}), respectively. We shall henceforth focus primarily on the
two solutions (\ref{QGKMI}) and (\ref{QGKMII}), to which we refer by I and II, respectively.
These solutions satisfy the regularity property (\ref{Kregular}), with
\be
\kappa= \left\{ \begin{array}{cl}
-2 e^{n \eta}\,\cosh( n \eta) &  \mbox{ for  } D_{n+1}^{(2)} - \text{I}\\
2 e^{n \eta}\,\sinh( n \eta)  &  \mbox{ for  } D_{n+1}^{(2)} - 
\text{II} \\
\end{array} \right. \,.
\label{kappaD2}
\ee

We observe that the $A_{2n-1}^{(2)}$ $K^{-}$-matrices (\ref{KA2oddsetI})-(\ref{KA2oddsetII}) and
the $D_{n+1}^{(2)}$ $K^{-}$-matrices (\ref{KD2}) are symmetric
\be
\left(K^{-}(u)\right)^{t} = K^{-}(u) \,.
\label{Ksymmetric}
\ee 

\section{Transfer matrix and Hamiltonian}\label{sec:transfer}

The transfer matrix $t(u)$ for an integrable open quantum spin chain with $N$ 
sites, which acts on the quantum space ${\cal V}^{\otimes N}$, is given by \cite{Sklyanin:1988yz}
\be
t(u) = \tr_a K^{+}_a(u)\, T_a(u)\,  K^{-}_{a}(u)\, \hat T_a(u) \,, 
\label{transfer}
\ee
where the monodromy matrices are defined by
\be 
T_a(u) = R_{aN}(u)\ R_{a N-1}(u)\ \cdots R_{a1}(u) \,,  \qquad
\hat T_a(u) = R_{1a}(u)\ \cdots R_{N-1 a}(u)\ R_{Na}(u) \,,  
\label{monodromy}
\ee
and the trace in (\ref{transfer}) is over the auxiliary space, which 
we denote by $a$. 
By construction, the transfer matrix satisfies the 
fundamental commutativity property \cite{Sklyanin:1988yz}
\be
\left[ t(u) \,, t(v) \right] = 0 \hbox{   for all   } u \,, v \,.
\label{commutativity}
\ee
The transfer matrix also has crossing symmetry \cite{Mezincescu:1991ag}
\be
t(u) = t(-u-\rho) \,.
\label{crossingtransfer}
\ee
The corresponding integrable open-chain Hamiltonian is given (up to 
multiplicative and additive constants) by $t'(0)$, which is local as 
a consequence of the regularity properties (\ref{Rregular}) and (\ref{Kregular}).
After some computation, one finds (up to terms proportional to the identity) \cite{Sklyanin:1988yz}
\be
{\cal H} \sim \sum_{k=1}^{N-1} h_{k,k+1} +\frac{1}{2\kappa} K^{-\, 
'}_{1}(0) + 
\frac{1}{\tr K^{+}(0)}\tr_{0} K^{+}_{0}(0) h_{N0}\,,
\label{Hamiltonian}
\ee
where the two-site Hamiltonian $h_{k,k+1}$ is given by
\be
h_{k,k+1} = \frac{1}{\xi(0)} {\cal P}_{k, k+1} R'_{k,k+1}(0)  \,.
\label{twosite}
\ee

We now proceed as in \cite{Ahmed:2017mqq} to simplify the expression
(\ref{Hamiltonian}), which will be needed in the subsequent section for proving the quantum
group invariance of the Hamiltonian.  Such a simplification is
possible due to the identity \cite{Ahmed:2017mqq}
\be
\tr_{1} K^{-}_{1}(-u-\rho)\, M_{1}\, R_{12}(2u)\, {\cal 
P}_{12}  = f(u)\, V_{2}\, K^{-\, t_{2}}_{2}(u)\, V_{2} \,,
\label{identity}
\ee
where $f(u)$ is a scalar function. In view of the isomorphism 
(\ref{isomorphism}) and the fact that the $K^{-}$-matrices are 
symmetric (\ref{Ksymmetric}), this identity can be rewritten as
\be
\tr_{1} K^{+}_{1}(u)\, {\cal P}_{12} R_{21}(2u)   = f(u)\, V_{2}\, 
K^{-}_{2}(u)\, V_{2} \,.
\label{identity2}
\ee
In order to go further, we must now consider the various cases 
separately.

\subsection{$A_{2n-1}^{(2)}$ - case I}

For the case $A_{2n-1}^{(2)} - \text{I}$ (\ref{KA2oddsetI}), it follows 
from (\ref{identity2}) as in \cite{Ahmed:2017mqq} that the third term in (\ref{Hamiltonian}) 
is proportional to the identity, while the second term evidently 
vanishes. The Hamiltonian therefore reduces to a sum of two-site Hamiltonians  \cite{Mezincescu:1990ui}
\be
{\cal H}^{(A_{2n-1}^{(2)} - \text{I})} = \sum_{k=1}^{N-1} h_{k,k+1} \,.
\label{HamiltonianA2oddI}
\ee
It is related to the transfer matrix (\ref{transfer}) by
\be
{\cal H}^{(A_{2n-1}^{(2)} - \text{I})} = \frac{1}{c_{1}} t'(0) + c_{2} \id^{\otimes N}\,,
\label{HA2oddIt}
\ee
with
\begin{align}
c_{1} &= 4^{N+1}  \sinh(2n\eta)\, 
\cosh(2(n+1)\eta)\, \sinh^{2N-1}(2\eta)\, \cosh^{2N}(2n\eta)\,, \non\\
c_{2} &=\frac{\cosh(2(3n+1)\eta)}{2 \sinh(4n\eta)\, 
\cosh(2(n+1)\eta)} \,.
\label{HA2Itc1c2}
\end{align}

\subsection{$A_{2n-1}^{(2)}$ - case II}

For the case $A_{2n-1}^{(2)} - \text{II}$ (\ref{KA2oddsetII}), we note
as in \cite{Ahmed:2017mqq} that (\ref{identity2}) implies
\begin{align}
f(0) &=  \frac{1}{\kappa}\, \xi(0) \tr K^{+}(0) \,,     \non \\
2 \tr_{1} K^{+}_{1}(0)\, {\cal P}_{12} R'_{21}(0) + \ldots &= f(0)\, V_{2}\,
K^{-\, '}_{2}(0) V_{2}\, + \ldots \,,
\label{intermed}
\end{align}
where the ellipses represent terms that are proportional to the 
identity, which we drop. We further note that 
\be
V\,  K^{-\, '}(0)\,  
V = - K^{-\, '}(0) \,.
\label{observe1}
\ee
It follows from (\ref{intermed}) and (\ref{observe1}) that 
\be
\frac{1}{\xi(0)\, \tr K^{+}(0)} \tr_{1} K^{+}_{1}(0)\, {\cal P}_{12} R'_{21}(0)
 = -\frac{1}{2\kappa}K^{-\, '}_{2}(0)  + \ldots
\ee
Let us define a new two-site Hamiltonian $\tilde h_{k,k+1}$ as in \cite{Ahmed:2017mqq}
\be
\tilde h_{k,k+1} \equiv h_{k,k+1} + \frac{1}{2\kappa} \left[ K^{-\, '}_{k}(0)
-  K^{-\, '}_{k+1}(0) \right] \,,
\label{twositeII}
\ee
where here $\kappa=1$ (\ref{kappaA2odd}). The Hamiltonian 
(\ref{Hamiltonian}) for this case is 
therefore again given by a sum of two-site Hamiltonians
\be
{\cal H}^{(A_{2n-1}^{(2)} - \text{II})} = \sum_{k=1}^{N-1} \tilde h_{k,k+1} \,.
\label{HamiltonianA2oddII}
\ee
This Hamiltonian is related to the transfer matrix (\ref{transfer}) by
\be
{\cal H}^{(A_{2n-1}^{(2)} - \text{II})} = \frac{1}{c_{1}} t'(0) + c_{2} \id^{\otimes N}\,,
\label{HA2oddIIt}
\ee
with
\begin{align}
c_{1} &= -4^{N+1}  \sinh(2n\eta)\, 
\cosh(2(n-1)\eta)\, \sinh^{2N-1}(2\eta)\, \cosh^{2N}(2n\eta)\,, \non\\
c_{2} &=\frac{\cosh(2(3n-1)\eta)}{2 \sinh(4n\eta)\, 
\cosh(2(n-1)\eta)} \,.
\label{HA2IItc1c2}
\end{align}

\subsection{$D_{n+1}^{(2)}$ - case I}

For the case $D_{n+1}^{(2)} - \text{I}$ (\ref{QGKMI}), we note that
\be
V\,  K^{-\, '}(0)\,  
V = - K^{-\, '}(0)  + \mu U +  \nu \id \,,
\label{observe}
\ee
where the matrix $U$ is defined as
\be
U=e_{n+1,n+1} + e_{n+1,n+2} + e_{n+2,n+1} + e_{n+2,n+2}\,,
\label{Umat}
\ee
and
\be
\mu =  -4 e^{n \eta}\sinh( n \eta)\,, \qquad
\nu = -2(3+e^{2 n \eta}) \,.
\ee 
It now follows from (\ref{intermed}) and (\ref{observe}) that 
\be
\frac{1}{\xi(0)\, \tr K^{+}(0)} \tr_{1} K^{+}_{1}(0)\, {\cal P}_{12} R'_{21}(0)
 = -\frac{1}{2\kappa}K^{-\, '}_{2}(0)  + \frac{\mu}{2\kappa} U_{2} + \ldots
\ee
The Hamiltonian (\ref{Hamiltonian}) for this case is therefore again given (up to a term proportional to 
$U_{N}$) by a sum of two-site Hamiltonians
\be
{\cal H}^{(D_{n+1}^{(2)} - \text{I})} = \sum_{k=1}^{N-1} \tilde h_{k,k+1} + 
\frac{\mu}{2\kappa} U_{N} \,,
\label{HamiltonianD2I}
\ee
where the two-site Hamiltonian is again given by (\ref{twositeII}), 
and $\kappa$ is now given by (\ref{kappaD2}).
This Hamiltonian is related to the transfer matrix (\ref{transfer}) by
\be
{\cal H}^{(D_{n+1}^{(2)} - \text{I})} = \frac{1}{c_{1}} t'(0) + c_{2} \id^{\otimes N}\,,
\label{HD2It}
\ee
with
\begin{align}
    c_{1} &=2^{4N+4} e^{6 n \eta} \left[\sinh(2 n \eta) 
    \sinh(2\eta)\right]^{2N-1} \non \\
    & \qquad \times \sinh((n+1)\eta) \sinh(4 n \eta) 
    \cosh^{2}(n \eta) \cosh((n-1)\eta)
    \,, \non \\
    c_{2} &= \frac{1}{2}\Big[\coth(\eta)-2\coth(2\eta)+2\coth(4 n 
    \eta)  + \coth((n+1)\eta) \non\\
    & \qquad + \tanh(\eta) + \tanh((n-1)\eta)+2\tanh(n 
    \eta) \Big]
    \,.
\label{HD2Itc1c2}    
\end{align}

\subsection{$D_{n+1}^{(2)}$ - case II}

For the case $D_{n+1}^{(2)} - \text{II}$  (\ref{QGKMII}), the relation (\ref{observe}) is again satisfied, with 
\be
\mu =  4 e^{n \eta}\cosh( n \eta)\,, \qquad
\nu = 2(-3+e^{2 n \eta}) \,.
\ee 
Hence, we similarly obtain
\be
{\cal H}^{(D_{n+1}^{(2)} - \text{II})} = \sum_{k=1}^{N-1} \tilde h_{k,k+1} + 
\frac{\mu}{2\kappa} U_{N} \,.
\label{HamiltonianD2II}
\ee

For $n=1$, it happens that $\tr K^{+}(0) = 0$. Hence, the Hamiltonian 
is related to the {\em second} derivative of the transfer matrix
\be
{\cal H}^{(D_{2}^{(2)} - II)} = \frac{1}{c_{1}} t''(0) + c_{2} 
\id^{\otimes N}\,, \qquad (n=1)
\label{HD2II1t}
\ee
with
\begin{align}
c_{1} &= 2^{4N+4} e^{6\eta} \sinh^{2}(\eta)\, \sinh^{2}(4\eta)\, \sinh^{4N-3}(2\eta)\,, \non\\
c_{2} &= \tanh(2\eta) +\frac{1}{4}\tanh(\eta) +\frac{5}{4}\coth(\eta)\,.
\label{HD2II1tc1c2}
\end{align}
For $n>1$, we find 
\be
{\cal H}^{(D_{n+1}^{(2)} - \text{II})} = \frac{1}{c_{1}} t'(0) + c_{2} 
\id^{\otimes N}\,, \qquad (n>1)
\label{HD2IIt}
\ee
with
\begin{align}
    c_{1} &=-2^{4N+4} e^{6 n \eta} \left[\sinh(2 n \eta) 
    \sinh(2\eta)\right]^{2N-1} \non \\
    & \qquad \times \sinh((n-1)\eta) \sinh(4 n \eta) 
    \sinh^{2}(n \eta) \cosh((n+1)\eta)
    \,, \non \\
    c_{2} &= \frac{1}{2}\Big[2\coth(4 n \eta) + \coth((n-1)\eta) + 
    2\coth(n\eta)  + \tanh((n+1)\eta) \Big]
    \,.
\end{align}

\section{Quantum group symmetries}\label{sec:QG}

We now show that the Hamiltonians constructed in the 
previous section have quantum group symmetry, and we discuss the 
consequences of this symmetry for the degeneracies and multiplicities 
of the spectrum. We consider each case separately.

\subsection{$A_{2n-1}^{(2)}$ - case I: $U_{q}(C_{n})$ symmetry}

A general argument was given in \cite{Mezincescu:1990ui} that the
Hamiltonian (\ref{HamiltonianA2oddI}) for the case $A_{2n-1}^{(2)} - \text{I}$
(\ref{KA2oddsetI}) has $U_{q}(C_{n})$ symmetry. Here we give a 
more explicit proof, by constructing the coproduct of the generators  
and showing that they commute with the Hamiltonian.

The vector representation of $C_{n} = Sp(2n)$ has dimension $2n$. In the orthogonal 
basis, the Cartan generators are given by
\be
H_{i} = e_{i, i} - e_{2n+1-i, 2n+1-i}\,, \qquad i = 1, \ldots, n \,,
\label{Cngens1}
\ee
while the generators corresponding to the simple roots are given by
\be 
E^{+}_{i} = \left\{
\begin{array}{ll}
    e_{i,i+1}  +  e_{2n-i, 2n+1-i} & i = 
    1,\ldots, n-1  \\
    \sqrt{2} e_{n, n+1} & i = n
    \end{array} \right. \,,
\label{Cngens2}
\ee
and $E^{-}_{i} = \left(E^{+}_{i}\right)^{t}$, where 
$e_{i j}$ are the elementary $(2n) \times (2n)$ 
matrices. These generators satisfy the commutation relations
\begin{align}
\left[H_i,E_j^\pm\right] &= \pm \alpha_i^{(j)}E_j^\pm \,, 
\label{HEcommutator}\\
\left[E_i^+,E_j^-\right] &=\delta_{i,j}\sum_{k=1}^n\alpha_k^{(j)}H_k \,,
\label{EEcommutator}
\end{align}
where $\{\alpha^{(1)}, \ldots, \alpha^{(n)}\}$ are the simple roots 
of $C_{n}$ in the orthogonal basis
\be
\alpha^{(j)} =  \left\{
\begin{array}{ll} e_{j} - e_{j+1} & j = 
    1,\ldots, n-1  \\
    2e_{n} & j = n
    \end{array} \right. \,,
\label{simplerootsCn}
\ee 
and $e_{j}$ are the elementary $n$-dimensional basis vectors 
$(e_{j})_{i} = \delta_{i,j}$ (i.e., $e_{1}=(1, 0, 0, 
\ldots, 0)\,, e_{2} = (0, 1, 0, \ldots, 0)$, etc.). 

We define the coproduct for these generators by
\begin{align}
\Delta(H_{j}) &= H_{j} \otimes \id + \id \otimes H_{j} \,, \qquad\qquad\qquad j = 
1, \ldots,  n \,, \non \\
\Delta(E^{\pm}_{j}) &= \left\{
\begin{array}{ll}
E^{\pm}_{j} \otimes e^{i \pi H_{j}} e^{\eta 
(H_{j} - H_{j+1})} + e^{-i \pi H_{j}} e^{-\eta 
(H_{j} - H_{j+1})} \otimes E^{\pm}_{j} & j = 1, \ldots, n-1 \\
E^{\pm}_{n} \otimes e^{2 \eta H_{n}} + e^{-2 \eta H_{n}} \otimes 
E^{\pm}_{n} & j=n
\end{array} \right. \,,
\label{coproductCn}
\end{align}
with $\Delta(\id)=\id\otimes \id$.
We note that
\be
\left[  \Delta(H_{i}) \,,   \Delta(E^{\pm}_{j}) \right] = \pm \alpha_{i}^{(j)} 
\Delta(E^{\pm}_{j})\,, \qquad i, j = 1, \ldots, n   
\label{DeltaHDeltaECn}
\ee
and 
\be
\Omega_{ij}\Delta(E_i^{+})\Delta(E_j^{-})-\Delta(E_j^{-})\Delta(E_i^{+})\Omega_{ij}
=\begin{cases}
\delta_{i,j}\frac{q^{\Delta(H_i)-\Delta(H_{i+1})}-q^{-\Delta(H_i)+\Delta(H_{i+1})}}
{q-q^{-1}} & i \mbox{ or } j \neq n\\
2\frac{q^{2\Delta(H_n)}-q^{-2\Delta(H_n)}}{q^{2}-q^{-2}} & i=j= n
\end{cases}\,,
\label{DeltaEDeltaECn}
\ee
where $q=e^{2\eta}$ and
\be 
\Omega_{ij}=\begin{cases}
e^{i \pi H_{\text{max}(i,j)}}\otimes\id & |i-j|=1 \mbox{ and } 1 \le 
\text{min}(i,j) \le n-2 \\
\quad \, \quad \id\otimes\id & \mbox{ otherwise }
\end{cases} \,.
\ee

By construction, the coproducts (\ref{coproductCn}) commute with the 
two-site Hamiltonian (\ref{twosite}),
\be
\left[ \Delta(H_{j}) \,, h_{1,2}  \right] = \left[ \Delta(E^{\pm}_{j}) 
\,, h_{1,2} \right] = 0 \,, \qquad j = 1, \ldots, n \,.
\label{QGtwosites}
\ee 
Since the $N$-site Hamiltonian (\ref{HamiltonianA2oddI}) is given by the 
sum of two-site Hamiltonians, it follows that 
the $N$-site Hamiltonian commutes with the $N$-fold 
coproducts\footnote{In more detail: let ${\cal H}_{(N)}$ denote the 
$N$-site Hamiltonian
so that ${\cal H}_{(2)} = h_{1,2}$. 
For $N=3$, according to (\ref{HamiltonianA2oddI}), we have
\be
{\cal H}_{(3)} = {\cal H}_{(2)} \otimes \id + \id \otimes {\cal H}_{(2)} 
\ee
and therefore
\be
\left[ \Delta_{(3)}(J) \,, {\cal H}_{(3)} \right] = 
\left[ \Delta_{(3)}(J) \,, {\cal H}_{(2)} \otimes \id \right] + 
\left[ \Delta_{(3)}(J) \,, \id \otimes {\cal H}_{(2)}  \right] \,,
\label{expl}
\ee
where $J$ is any one of the generators $H_{j}$ or $E^{\pm}_{j}$. 
The coproduct (\ref{coproductCn}) is coassociative, i.e.
\be
(\Delta \otimes \id) \Delta = (\id \otimes \Delta) \Delta \,.
\ee
Thus, setting
\be
\Delta(J) = \sum_{i} a_{i} \otimes b_{i} \,,
\ee
we see that
\be
\Delta_{(3)}(J) = \sum_{i} \Delta(a_{i}) \otimes b_{i} = \sum_{i}  a_{i}\otimes \Delta(b_{i}) \,.
\label{Delta3J}
\ee
It follows from (\ref{QGtwosites}) and (\ref{Delta3J})
that the commutators in (\ref{expl}) separately vanish; 
and similarly for higher $N$.}
\be
\left[ \Delta_{(N)}(H_{j}) \,, {\cal H}^{(A_{2n-1}^{(2)} - \text{I})} \right] = \left[ \Delta_{(N)}(E^{\pm}_{j}) 
\,, {\cal H}^{(A_{2n-1}^{(2)} - \text{I})} \right] = 0 \,, \qquad j = 1, \ldots, n \,.
\label{UqCnsymmetry}
\ee 
This provides an explicit demonstration of the $U_{q}(C_{n})$ 
invariance of the Hamiltonian ${\cal H}^{(A_{2n-1}^{(2)} - \text{I})}$.

The transfer matrix $t(u)$ (\ref{transfer}) also has this symmetry 
\cite{Mezincescu:1991rb}, so in particular it commutes with the 
Cartan generators
\be
\left[ \Delta_{(N)}(H_{j}) \,, t(u)^{(A_{2n-1}^{(2)} - \text{I})} \right]  = 0 \,, 
\qquad j = 1, \ldots, n \,.
\label{transfCartanCn}
\ee 

\subsubsection{Degeneracies and multiplicities for $U_{q}(C_{n})$}

The symmetry (\ref{UqCnsymmetry}) implies that the eigenstates of the 
Hamiltonian form irreducible representations of $U_{q}(C_{n})$. For generic 
values of $\eta$, the representations are the same as for the 
classical algebra $C_{n}$. The $N$-site Hilbert space can therefore 
be decomposed into a direct sum of irreducible representations of $C_{n}$
\be
{\cal V}^{(2n) \otimes N} = \bigoplus_{j} d^{(j, N, n)}\, {\cal V}^{(j)} \,,
\label{decomCn}
\ee 
where ${\cal V}^{(j)}$ denotes an irreducible representation of
$C_{n}$ with dimension $j$ (= degeneracy of the corresponding energy
eigenvalue) and $d^{(j, N, n)}$ is its multiplicity. 

We present the first few cases below, denoting the irreducible 
representations of $C_{n}$ both by their dimensions (in boldface) and by their 
Dynkin labels $[a_{1}, \ldots, a_{n}]$ (see e.g. \cite{Feger:2012bs}): 
\begin{align*}
C_1 = A_{1}: \qquad  N=2: \qquad \bf{2}\otimes\bf{2}& 
=\bf{1}\oplus \bf{3}\\
&=[0]\oplus[2]
\end{align*}
\begin{align}
\qquad\qquad\qquad\qquad N=3: \qquad \bf{2}\otimes \bf{2}\otimes 
\bf{2}&=2\cdot\bf{2}\oplus \bf{4}\nonumber\\
&=2[1]\oplus [3]
\label{decompC1}
\end{align}
\begin{align*}    
C_2: \qquad  N=2: \qquad \bf{4}\otimes\bf{4}& =\bf{1}\oplus \bf{5}\oplus\bf{10}\\
&=[0,0]\oplus[0,1]\oplus[2,0]
\end{align*}
\begin{align}
\qquad\qquad\qquad N=3: \qquad \bf{4}\otimes \bf{4}\otimes 
\bf{4}&=3\cdot\bf{4}\oplus \mbox{2}\cdot\bf{16}\oplus \bf{20}\nonumber\\
&=3[1,0]\oplus 2[1,1]\oplus[3,0]
\label{decompC2}
\end{align}
\begin{align*}
C_3: \qquad  N=2: \qquad \bf{6}\otimes\bf{6}&=\bf{1}\oplus \bf{14}\oplus\bf{21}\\
&=[0,0,0]\oplus[0,1,0]\oplus[2,0,0]
\end{align*}
\begin{align}
\qquad\qquad\qquad\qquad\qquad N=3: \qquad \bf{6}\otimes \bf{6}\otimes \bf{6}&=
3\cdot\bf{6}\oplus\bf{14'}\oplus \bf{56}\oplus \mbox{2}\cdot\bf{64}\nonumber\\
	&=3[1,0,0]\oplus [0,0,1]\oplus[3,0,0]\oplus 2[1,1,0]
\label{decompC3}
\end{align}
We have verified numerically that the Hamiltonian as well as the
transfer matrix for the case $A_{2n-1}^{(2)} - \text{I}$ (\ref{KA2oddsetI})
have exactly these degeneracies and multiplicities for generic values
of $\eta$, which provides further evidence of their $U_{q}(C_{n})$
invariance.

\subsection{$A_{2n-1}^{(2)}$ - case II: $U_{q}(D_{n})$ symmetry}     

As noted in the Introduction, we expect that the
Hamiltonian (\ref{HamiltonianA2oddII}) for the case $A_{2n-1}^{(2)} 
- \text{II}$
(\ref{KA2oddsetII}) with $n>1$ has $U_{q}(D_{n})$ symmetry. We now 
proceed to explicitly demonstrate this symmetry.

The vector representation of $D_{n} = O(2n)$ has dimension $2n$. 
In the orthogonal basis, the Cartan generators are given by
\be
H_{i} = e_{i, i} - e_{2n+1-i, 2n+1-i}\,, \qquad i = 1, \ldots, n \,,
\label{Dngens1}
\ee
and the generators corresponding to the simple roots are given by
\be 
E^{+}_{i} = \left\{
\begin{array}{ll}
    e_{i,i+1}  +  e_{2n-i, 2n+1-i} & i = 
    1,\ldots, n-1  \\
    e_{n-1, n+1} + e_{n, n+2} & i = n
    \end{array} \right. \,,
\label{Dngens2}
\ee
with
$E^{-}_{i} = \left(E^{+}_{i}\right)^{t}$. They are the same as the $C_{n}$ 
generators (\ref{Cngens1}) and (\ref{Cngens2}), except for 
$E^{\pm}_{n}$. These generators satisfy the commutation relations (\ref{HEcommutator}) and 
(\ref{EEcommutator}), where
$\{\alpha^{(1)}, \ldots, \alpha^{(n)}\}$ are now
the simple roots of $D_{n}$ in the orthogonal basis, which are given by
\be
\alpha^{(j)} =  \left\{
\begin{array}{ll} e_{j} - e_{j+1} & j = 
    1,\ldots, n-1  \\
    e_{n-1} + e_{n} & j = n
    \end{array} \right. \,,
\label{simplerootsDn}
\ee 
c.f. (\ref{simplerootsCn}). 
It useful to introduce an additional pair of generators 
\begin{equation}
E^{+}_0=e_{1,2n-1}+e_{2,2n} \,, \qquad  
E^{-}_0=\left(E^{+}_0\right)^t \,,
\end{equation}
\noindent
which are related to $E_n^{\pm}$ as follows
\begin{equation}
E_n^{\pm}=\begin{cases}
\hfil E_0^{\pm} & n=2\\
\hfil -\left[\left[E_0^{\pm},E_2^{\mp}\right],E_1^{\mp}\right] & n=3\\
\hfil \left[\left[ \left[\left[E_0^{\pm},E_2^{\mp}\right],E_3^{\mp}\right],E_1^{\mp}\right],E_2^{\mp}\right] & n=4\\
\hfil -\left[\left[ \left[\left[\left[\left[E_0^{\pm},E_2^{\mp}\right],E_3^{\mp}\right],E_4^{\mp}\right],E_1^{\mp}\right],E_2^{\mp}\right],E_3^{\mp}\right] & n=5\\
\hfil ... & \\
\hfil (-1)^n\left[\left[\left[\left[\left[\left[...\left[E_0^{\pm},E_2^{\mp}\right],E_3^{\mp}\right],\,...\,,E_{n-1}^{\mp}\right],E_1^{\mp}\right],E_2^{\mp}\right],E_3^{\mp}\right],...,E_{n-2}^{\mp}\right] & \hfil n
\end{cases} 
\label{EnDn}
\end{equation}
where the final line has a $2(n-2)$-fold multiple commutator.

We define the coproduct for the Cartan generators and the first $n-1$ 
raising/lowering operators as before (\ref{coproductCn})
\begin{align}
\Delta(H_{j}) &= H_{j} \otimes \id + \id \otimes H_{j} \,, &  j &= 
1, \ldots,  n \,, \non \\
\Delta(E^{\pm}_{j}) &= 
E^{\pm}_{j} \otimes e^{i \pi H_{j}} e^{\eta 
(H_{j} - H_{j+1})} + e^{-i \pi H_{j}} e^{-\eta 
(H_{j} - H_{j+1})} \otimes E^{\pm}_{j} \,,   & j &= 1, \ldots, n-1 
\,,
\label{coproductDn}
\end{align}
with $\Delta(\id)=\id\otimes \id$.
Defining the coproduct for the additional generators $E_0^{\pm}$  by
\begin{equation}
\Delta(E_0^{\pm})=E_0^{\pm}\otimes e^{i \pi H_2}+e^{2\eta\left(H_1+H_2\right)+i\pi H_2}\otimes E_0^{\pm}
\label{DeltaE0}
\end{equation}
allows us to write the coproduct for $E_n^{\pm}$ using (\ref{EnDn}) 
and (\ref{coproductDn}) as
\begin{equation}
\Delta(E_n^{\pm})=(-1)^n\left[\left[\left[\left[\left[...\left[\Delta(E_0^{\pm}),\Delta(E_2^{\mp})\right],\Delta(E_3^{\mp})\right],...\Delta(E_{n-1}^{\mp})\right],\Delta(E_1^{\mp})\right],\Delta(E_2^{\mp})\right],...\Delta(E_{n-2}^{\mp})\right].
\label{DeltaEnDn}
\end{equation}
These coproducts satisfy (\ref{DeltaHDeltaECn}); and, for $i,j<n$, 
also (\ref{DeltaEDeltaECn}). 

By construction, the coproducts 
(\ref{coproductDn})-(\ref{DeltaEnDn})
commute with the ``new'' two-site Hamiltonian (\ref{twositeII})
\be
\left[ \Delta(H_{j}) \,, \tilde h_{1,2}  \right] = \left[ \Delta(E^{\pm}_{j}) 
\,, \tilde h_{1,2} \right] = 0 \,, \qquad j = 1, \ldots, n \,.
\ee 
Since the $N$-site Hamiltonian (\ref{HamiltonianA2oddII}) is given by 
a sum of such two-site Hamiltonians, the $N$-site Hamiltonian commutes with the $N$-fold coproducts
\be
\left[ \Delta_{(N)}(H_{j}) \,, {\cal H}^{(A_{2n-1}^{(2)} - \text{II})} \right] = \left[ \Delta_{(N)}(E^{\pm}_{j}) 
\,, {\cal H}^{(A_{2n-1}^{(2)} - \text{II})} \right] = 0 \,, \qquad j = 1, \ldots, n \,,
\label{UqDnsymmetry}
\ee 
which implies the $U_{q}(D_{n})$ invariance of the Hamiltonian ${\cal
H}^{(A_{2n-1}^{(2)} - \text{II})}$.  

The Cartan generators commute with the transfer matrix $t(u)$ (\ref{transfer})
\be
\left[ \Delta_{N}(H_{j}) \,, t(u)^{(A_{2n-1}^{(2)} - \text{II})} \right] = 0 \,, 
\qquad j = 1, \ldots, n \,.
\label{transfCartanDn}
\ee
We conjecture that the transfer matrix for the case $A_{2n-1}^{(2)} 
- \text{II}$ (\ref{KA2oddsetII}) with $n>1$ is in fact $U_{q}(D_{n})$ invariant.

\subsubsection{Degeneracies and multiplicities for 
$U_{q}(D_{n})$}\label{sec:degmultDn}

The symmetry (\ref{UqDnsymmetry}) implies that the eigenstates of the 
Hamiltonian form irreducible representations of $U_{q}(D_{n})$. For generic 
values of $\eta$, the representations are the same as for the 
classical algebra $D_{n}$. The $N$-site Hilbert space can therefore 
be decomposed into a direct sum of irreducible representations of 
$D_{n}$, similarly to (\ref{decomCn}).

The first few cases are as follows: 
\begin{align*}    
D_2: \qquad  N=2: \qquad \bf{4}\otimes\bf{4}& 
=\bf{1}\oplus \bf{3}\oplus \overline{\bf{3}}\oplus\bf{9}\\
&=[0,0]\oplus[2,0]\oplus[0,2]\oplus[2,2]
\label{D2decomp}
\end{align*}
\begin{align}
\qquad\qquad\qquad\qquad\qquad N=3: \qquad \bf{4}\otimes \bf{4}\otimes 
\bf{4}&=4\cdot\bf{4}\oplus \mbox{2}\cdot\bf{8}\oplus 
\mbox{2}\cdot\overline{\bf{8}}\oplus \bf{16}\nonumber\\
&=4[1,1]\oplus 2[3,1]\oplus 2[1,3]\oplus [3,3]
\end{align}
\begin{align*}
D_3: \qquad  N=2: \qquad \bf{6}\otimes\bf{6}&=\bf{1}\oplus 
\bf{15}\oplus\bf{20'}\\
&=[0,0,0]\oplus[0,1,1]\oplus[2,0,0]
\end{align*}
\begin{align}
\qquad\qquad\qquad\qquad\qquad\quad\ N=3: \qquad \bf{6}\otimes \bf{6}\otimes \bf{6}&=
3\cdot\bf{6}\oplus\bf{10}\oplus\overline{\bf{10}}\oplus \bf{50}\oplus \mbox{2}\cdot\bf{64}\nonumber\\
	&=3[1,0,0]\oplus [0,2,0]\oplus [0,0,2]\oplus[3,0,0]\oplus 
	2[1,1,1]
\label{D3decomp}
\end{align}

Surprisingly, the Hamiltonian and the transfer matrix for the case $A_{2n-1}^{(2)} - \text{II}$ (\ref{KA2oddsetII})
do {\em not} have precisely these degeneracies and multiplicities for generic 
values of $\eta$. Indeed, we observe that their degeneracies are {\em higher}:

\begin{align}
n=2: \qquad\qquad & N=2: \qquad \{ 1\,, 6\,, 9 \}\non \\
& N=3: \qquad \{4\cdot 4\,, 3\cdot 16 \}
\label{D2degen}
\end{align}
\begin{align}
n=3: \qquad\qquad & N=2: \qquad \{1\,, 15\,, 20 \}\non \\
& N=3: \qquad \{3\cdot 6\,, 20\,, 50\,, 2\cdot 64 \}
\label{D3degen}
\end{align}

\noindent
Comparing (\ref{D2decomp}) and  (\ref{D3decomp}) with (\ref{D2degen}) 
and (\ref{D3degen}) respectively, we see that 
the observed degeneracies would be explained if the Hamiltonian 
and the transfer matrix have an 
additional $Z_{2}$ symmetry that maps $D_{n}$ representations to 
their conjugates, which would imply that
representations and their conjugates (for 
example, $\bf{3}$ and $\overline{\bf{3}}$) are degenerate. We have 
explicitly constructed such symmetry transformations for small values 
of $n$ and $N$. We conjecture that such symmetry transformations exist for general 
values of $n$ and $N$.

\subsection{$D_{n+1}^{(2)}$ - case I: $U_{q}(B_{n})$ symmetry}

As discussed in the Introduction, we expect that the
Hamiltonian (\ref{HamiltonianD2I}) for the case $D_{n+1}^{(2)} - \text{I}$
(\ref{QGKMI}) has $U_{q}(B_{n})$ symmetry. We now proceed to 
explicitly demonstrate this symmetry.

The first step to demonstrating this symmetry is to construct the generators of $B_{n}$. Although the vector 
representation of $B_{n}$ has dimension $2n+1$, here we need 
generators that act on a vector space whose dimension is one greater, 
i.e. $d=2n+2$. An 
appropriate embedding can be found by studying the symmetries of the 
transfer matrix with one site ($N=1$). Hence, we choose the Cartan generators $H_j$ 
\be
H_j=e_{j,j}-e_{2n+3-j,2n+3-j}\,,\quad j =1,\dots,n \,,
\label{CartanBn}
\ee
and the generators $E_j^\pm$ corresponding to simple roots 
\be
E_j^+=
\begin{cases}
e_{j,j+1}+e_{2n+2-j,2n+3-j}\,,\qquad\qquad\qquad\quad 1\leq j\leq n-1\,, \\
\frac{1}{\sqrt{2}}\left(e_{n,n+1}-e_{n,n+2}+e_{n+2,n+3}-e_{n+1,n+3}\right)\,,\quad j=n\,,
\end{cases}
\label{EpBn}
\ee
with $E_j^- = \left(E_j^+\right)^{t}$, where $e_{i,j}$ are the $(2n+2)\times(2n+2)$ elementary matrices.
These generators satisfy the commutation relations (\ref{HEcommutator}) and 
(\ref{EEcommutator}), where $\{ \alpha^{(j)} \}$ are the simple roots of $B_{n}$ in the orthogonal basis
\be
\alpha^{(j)} =  \left\{
\begin{array}{ll} e_{j} - e_{j+1} & j = 
    1,\ldots, n-1  \\
    e_{n} & j = n
    \end{array} \right. \,,
\label{simplerootsBn}
\ee 
c.f. (\ref{simplerootsCn}), (\ref{simplerootsDn}).
We also introduce the generators $E_{0}^{\pm}$ defined by
\be
E_{0}^{+} = 
\frac{1}{\sqrt{2}}\left[e_{1,n+1}-e_{1,n+2}+(-1)^{n}\left(e_{n+1,2n+2}-e_{n+2,2n+2}\right) \right]\,, 
\qquad E_{0}^{-} = (E_{0}^{+})^{t} \,,
\ee
which are related to $E_{n}^{\pm}$ by $(n-1)$-fold 
multiple commutators as follows 
\be
E^{+}_{n} =\left\{
\begin{array}{cl}
E^{+}_{0} & n=1 \\
-[ E^{+}_{0}\,, E^{-}_{1}] & n=2 \\
\ [[ E^{+}_{0}\,, E^{-}_{1}]\,, E^{-}_{2}] & n=3 \\
\vdots & \quad\vdots \\
(-1)^{n+1} [\ldots [ E^{+}_{0}\,, E_{1}^{-}]\,, \ldots \,,
E_{n-1}^{-}] & \quad n 
\end{array} \right. \,, \qquad E_{n}^{-} = (E_{n}^{+})^{t} \,.
\label{Enested}
\ee

We propose the following expressions for the coproduct:
\begin{align}
\Delta(H_j) &=H_j\otimes \id+\id\otimes 
H_j\,,\qquad\qquad\qquad\qquad\qquad\qquad j=1,\dots, n\,, \non \\
\Delta(E_j^\pm) &=E_j^\pm\otimes e^{i\pi 
H_{j+1}}+e^{-2\eta(H_j-H_{j+1})+i\pi H_{j+1}}\otimes E_j^\pm\,,\quad 
j=1,\dots,n-1 \,,
\label{coproductBn}
\end{align}
with $\Delta(\id)=\id\otimes \id$.
Moreover, using the result (\ref{Enested}) together with
\be
\Delta(E_0^\pm)=
\begin{cases}
E_0^\pm\otimes e^{-\eta H_{1}}+e^{\eta H_{1}}\otimes E_0^\pm\,, 
\qquad\qquad\qquad\qquad n = \mbox{even} \,, \\
E_0^\pm\otimes e^{(i\pi-\eta) H_{1}}+e^{-(i\pi-\eta) 
H_{1}}\otimes E_0^\pm\,, \qquad\qquad\ n = \mbox{odd} \,.
\end{cases}
\ee
we obtain
\be
\Delta(E_n^\pm) = (\mp 1)^{n+1} [\ldots [ \Delta(E^{\pm}_{0})\,, 
\Delta(E_{1}^{\mp})]\,, \ldots \,,
\Delta(E_{n-1}^{\mp})] \,.
\label{coproductEnBn}
\ee

The above coproduct satisfies
\begin{align}
\Delta(E_j^+)\Delta(E_j^-)-e^{4\eta}\Delta(E_j^-)\Delta(E_j^+) &=
\frac{e^{-4\eta(\Delta(H_j)-\Delta(H_{j+1}))}-\id\otimes\id}{e^{-4\eta}-1}\,,\quad
j=1,\dots,n-1 \,, \\
\Delta(E_0^+)\Delta(E_0^-)-\Delta(E_0^-)\Delta(E_0^+) &=
%H_1\otimes e^{-2\eta H_1}+e^{2\eta H_1}\otimes H_1 
\frac{e^{2\eta\Delta(H_1)}-e^{-2\eta\Delta(H_1)}}{e^{2\eta}-e^{-2\eta}} \,.
\end{align}

By construction, these coproducts commute with the two-site 
Hamiltonian (\ref{twositeII})
\be
\left[ \Delta(H_{j}) \,, \tilde h_{1,2}  \right] = \left[ \Delta(E^{\pm}_{j}) 
\,, \tilde h_{1,2} \right] = 0 \,, \qquad j = 1, \ldots, n \,.
\ee 
All the generators also commute with the matrix $U$ (\ref{Umat})
\be
\left[H_{j} \,, U \right] = \left[ E^{\pm}_{j} \,, U \right]  = 0 \,, \qquad j = 1, \ldots, n \,.
\ee
It follows that 
the $N$-site Hamiltonian (\ref{HamiltonianD2I}) commutes with the $N$-fold coproducts
\be
\left[ \Delta_{(N)}(H_{j}) \,, {\cal H}^{(D_{n+1}^{(2)} - \text{I})} \right] = \left[ \Delta_{(N)}(E^{\pm}_{j}) 
\,, {\cal H}^{(D_{n+1}^{(2)} - \text{I})} \right] = 0 \,, \qquad j = 1, \ldots, n \,,
\label{UqBnsymmetry}
\ee 
which implies the $U_{q}(B_{n})$ invariance of the Hamiltonian ${\cal H}^{(D_{n+1}^{(2)} - \text{I})}$.

It is not difficult to show that the Cartan generators also commute with the transfer matrix 
\be
\left[ \Delta_{N}(H_{j}) \,, t(u)^{(D_{n+1}^{(2)} - \text{I})} \right] = 0 \,, \qquad j = 1, \ldots, n \,.
\label{transfCartanBn}
\ee
We conjecture that the transfer matrix for the case $D_{n+1}^{(2)} - 
\text{I}$ (\ref{QGKMI}) is in fact $U_{q}(B_{n})$ invariant.

\subsubsection{Degeneracies and multiplicities for $U_{q}(B_{n})$}\label{sec:degmult}

The $U_{q}(B_{n})$ invariance (\ref{UqBnsymmetry}) of the Hamiltonian implies that,
for generic values of $\eta$,  the $N$-site Hilbert space 
has a decomposition of the following form
\be
\left( {\cal V}^{(2n+1)} \oplus {\cal V}^{(1)} \right)^{\otimes N} = 
\bigoplus_{j} d^{(j, N, n)}\, {\cal V}^{(j)} \,,
\label{decomBn}
\ee
where ${\cal V}^{(j)}$ denotes an irreducible representation of $B_{n}$
with dimension $j$ (= degeneracy of the corresponding 
energy eigenvalue) and $d^{(j, N, n)}$ is its multiplicity.

We present the first few cases below, again denoting the irreducible 
representations both by their dimensions (in boldface) and by their 
Dynkin labels $[a_{1}, \ldots, a_{n}]$:
\begin{align}
B_{1} = A_{1}:\qquad  N&=2: & \left({\bf 3} \oplus {\bf 1}\right)^{\otimes 2} 
&= 2\cdot {\bf 1} \oplus  3 \cdot{\bf 3} 
\oplus  {\bf 5}  \non \\
&  & &= 2 [0] \oplus 3 [2] \oplus [4] \non\\
 N&=3: & \left({\bf 3} \oplus {\bf 1}\right)^{\otimes 3} 
 &= 5 \cdot {\bf 1} 
 \oplus  9 \cdot {\bf 3} 
\oplus 5 \cdot {\bf 5} \oplus {\bf 7}  \non\\
&   &  & = 5 [0] \oplus 9 [2] \oplus 5 [4] \oplus [6]
\label{decompB1}
\end{align}  
\begin{align}
B_{2}:\qquad  N&=2: & \left({\bf 5} \oplus {\bf 1}\right)^{\otimes 2} 
&= 2\cdot {\bf 1} \oplus  2 \cdot{\bf 5} 
\oplus  {\bf 10} \oplus  {\bf 14} \non \\
&  & &= 2 [0, 0] \oplus 2 [1, 0] \oplus [0, 
2]  \oplus [2, 0]\non\\
 N&=3: & \left({\bf 5} \oplus {\bf 1}\right)^{\otimes 3} 
 &= 4 \cdot {\bf 1} 
 \oplus  6 \cdot {\bf 5} 
\oplus 4 \cdot {\bf 10} \oplus  3\cdot {\bf 14}  \oplus {\bf 30}  \oplus  
2\cdot {\bf 35}\non\\
&   &  & = 4 [0, 0] \oplus 6 [1, 0] \oplus 4 [0, 2] 
\oplus 3 [2, 0] \oplus  [3, 0] \oplus 2 [1, 2]
\label{decompB2}
\end{align}  
\begin{align}
B_{3}:\qquad  N&=2: & \left({\bf 7} \oplus {\bf 1}\right)^{\otimes 2} 
&= 2\cdot {\bf 1} \oplus  2 \cdot{\bf 7} 
\oplus  {\bf 21} \oplus  {\bf 27} \non \\
&  & &= 2 [0,0,0] \oplus 2 [1,0,0] \oplus 
[0,1,0] \oplus [2,0,0] \non\\
 N&=3: & \left({\bf 7} \oplus {\bf 1}\right)^{\otimes 3} 
 &= 4 \cdot {\bf 1} 
 \oplus  6 \cdot {\bf 7} 
\oplus 3 \cdot {\bf 21}  \oplus 3 \cdot {\bf 27} \oplus {\bf 
35} \oplus {\bf 77}\oplus  2 \cdot {\bf 105}\non \\
& & &= 4 [0,0,0] \oplus 6 [1,0,0] \oplus 3 
[0,1,0] \oplus 3 [2,0,0] \oplus \non\\
&  & & \qquad \oplus  [0,0,2] \oplus [3,0,0] \oplus 2 
[1,1,0] 
\label{decompB3}
\end{align}  

As in the case $A_{2n-1}^{(2)} - \text{II}$ discussed in Section
\ref{sec:degmultDn}, the Hamiltonian and the transfer matrix for the
case $D_{n+1}^{(2)} - \text{I}$ (\ref{QGKMI}) do {\em not} have precisely
these degeneracies and multiplicities for generic values of $\eta$.
Indeed, we observe that their degeneracies are {\em higher}:

\begin{align}
n=1: \qquad\qquad & N=2: \qquad \{ 2\cdot 1\,, 3\,, 5\,, 6 \}\non \\
& N=3: \qquad \{3\cdot 1\,, 2\,, 3\cdot 3\,, 5\,, 3\cdot 6\,, 7\,, 
2\cdot 10 \}
\label{B1degen}
\end{align}

\begin{align}
n=2: \qquad\qquad & N=2: \qquad \{ 2\cdot 1\,, 2\cdot 5\,, 10\,, 14 \}\non \\
& N=3: \qquad \{ 4 \cdot 1\,,   6 \cdot 5\,,  2 \cdot 10\,,  3\cdot 
14\,,  20\,, 30\,,  2\cdot 35 \}
\label{B2degen}
\end{align}

\begin{align}
n=3: \qquad\qquad & N=2: \qquad \{ 2\cdot 1\,, 2\cdot 7\,, 21\,, 27 \}\non \\
& N=3: \qquad \{ 4 \cdot 1\,, 6 \cdot 7\,, 3 \cdot 21\,,   3 \cdot 
27\,,  35\,, 77\,,   2 \cdot 105 \}
\label{B3degen}
\end{align}

\noindent
Comparing the $n=2$ results (\ref{decompB2}) and (\ref{B2degen}), we see that 
the observed degeneracies are almost the same as those predicted from 
$U_{q}(B_{2})$ symmetry; the one exception occurs for $N=3$, where
two of the four ${\bf 10}$'s are degenerate, yielding a 20-fold degeneracy.
Moreover, comparing the $n=3$ results (\ref{decompB3}) 
and (\ref{B3degen}),  we see that 
the observed degeneracies are exactly the same as those predicted from 
$U_{q}(B_{3})$ symmetry.

However, comparing the $n=1$ results (\ref{decompB1}) 
and (\ref{B1degen}), we see that 
the observed degeneracies for many of the levels are higher than expected from 
$U_{q}(B_{1})$ symmetry: for $N=2$, two of the three ${\bf 3}$'s are 
degenerate; and for $N=3$, two pairs of ${\bf 5}$'s are 
degenerate, etc. It would be interesting to find a symmetry (such as 
the $Z_{2}$ symmetry proposed for the case $A_{2n-1}^{(2)} - \text{II}$)
that can account for these higher degeneracies.

\subsection{$D_{n+1}^{(2)}$ - case II: $U_{q}(B_{n})$ symmetry}

As discussed in the Introduction, we expect that the
Hamiltonian (\ref{HamiltonianD2II}) for the case $D_{n+1}^{(2)} - 
\text{II}$
(\ref{QGKMII}) also has $U_{q}(B_{n})$ symmetry. In fact, the 
argument is exactly the same as for the case $D_{n+1}^{(2)} - \text{I}$: 
the $B_{n}$ generators and their coproducts are the same (see Eqs. 
(\ref{CartanBn}), (\ref{EpBn}), (\ref{coproductBn}), 
(\ref{coproductEnBn})), and we similarly obtain
\be
\left[ \Delta_{(N)}(H_{j}) \,, {\cal H}^{(D_{n+1}^{(2)} - \text{II})} \right] = \left[ \Delta_{(N)}(E^{\pm}_{j}) 
\,, {\cal H}^{(D_{n+1}^{(2)} - \text{II})} \right] = 0 \,, \qquad j = 1, \ldots, n \,,
\label{UqBnsymmetryII}
\ee 
which implies the $U_{q}(B_{n})$ invariance of the Hamiltonian ${\cal 
H}^{(D_{n+1}^{(2)} - \text{II})}$.

Although ${\cal H}^{(D_{n+1}^{(2)} - \text{I})}$ and ${\cal 
H}^{(D_{n+1}^{(2)} - \text{II})}$ 
have different spectra, the degeneracies and multiplicities are the 
same.

\section{Bethe ansatz for $A_{2n-1}^{(2)}$ - cases I and II}\label{sec:BAA2odd}

For the case $A_{2n-1}^{(2)} - \text{I}$ (\ref{KA2oddsetI}), the 
eigenvalues of the transfer matrix (\ref{transfer}) have been determined by both analytical 
Bethe ansatz \cite{Artz:1995bm} and nested algebraic Bethe ansatz 
\cite{Li:2005pp}; however, for the case $A_{2n-1}^{(2)} - \text{II}$ 
(\ref{KA2oddsetII}), the eigenvalues have not (to our knowledge) been investigated until 
now. In this section, we recall the Bethe ansatz solution for the case 
$A_{2n-1}^{(2)} - \text{I}$, and we propose its generalization for the case 
$A_{2n-1}^{(2)} - \text{II}$. We also propose for both cases a formula for the 
Dynkin label $[a_{1}\,, \ldots\,, a_{n}]$ of a Bethe state in terms of 
the cardinalities $(m_{1}\,, \ldots\,, m_{n})$ of the corresponding 
Bethe roots, which determines the degeneracy of the corresponding 
eigenvalue. Finally, we check the completeness of the Bethe ansatz 
solutions numerically for small values of $n$ and $N$.

\subsection{Transfer matrix eigenvalues}

For real values of $\eta$, the transfer matrix for both cases
$A_{2n-1}^{(2)} - \text{I}$ and $A_{2n-1}^{(2)} - \text{II}$ 
is Hermitian.
The commutativity property (\ref{commutativity}) and the fact that the 
transfer matrix also commutes with all of the Cartan generators 
(\ref{transfCartanCn}), (\ref{transfCartanDn}) imply that the transfer matrix and Cartan 
generators can be simultaneously diagonalized
\begin{align}
t(u)\, |\Lambda^{(m_{1}, \ldots\,, m_{n})}\rangle &= \Lambda^{(m_{1}, 
\ldots, m_{n})}(u)\,
|\Lambda^{(m_{1}, \ldots, m_{n})}\rangle \,, \non \\
\Delta_{N}(H_{j})\, |\Lambda^{(m_{1}, \ldots, m_{n})}\rangle &= h_{j}\,
|\Lambda^{(m_{1}, \ldots, m_{n})}\rangle \,, \qquad j = 1, \ldots, n 
\,,
\end{align}
where the Bethe states $|\Lambda^{(m_{1}, \ldots, m_{n})}\rangle$ are 
independent of the spectral parameter. We assume that the Bethe 
states are highest-weight states of $U_{q}(C_{n})$ (case I) or 
$U_{q}(D_{n})$ (case II)
\be
\Delta_{N}(E^{+}_{j})\, |\Lambda^{(m_{1}, \ldots, m_{n})}\rangle =0 \,, 
\qquad j = 1, \ldots, n 
\,.
\label{highestweightA2odd}
\ee
The Bethe states depend on $n$ sets of Bethe roots $\{u^{(1)}_{1}, \ldots, 
u^{(1)}_{m_{1}}\}, \ldots, \{u^{(n)}_{1}, \ldots, 
u^{(n)}_{m_{n}}\}$, with cardinalities $m_{1}\,, \ldots\,, m_{n}$, 
respectively,
\be
|\Lambda^{(m_{1}, \ldots, m_{n})} \rangle = |\{u^{(1)}_{1}, \ldots, 
u^{(1)}_{m_{1}}\}, \ldots, \{u^{(n)}_{1}, \ldots, 
u^{(n)}_{m_{n}}\} \rangle \,.
\label{Bethestates}
\ee

The eigenvalues of the transfer matrix are given by
\begin{IEEEeqnarray}{rCl}
& & \Lambda^{(m_1 \,, \cdots \,, m_n)}(u) \non \\
& &  =  A^{(m_1)}(u)\, \psi(u)\
\frac{\sinh(u-4n\eta)}{\sinh(u-2\eta)}
\frac{\cosh(u-2(n+1)\eta)}{\cosh(u-2n\eta)} 
\left[ 2 \sinh (\uh -2\eta) \cosh(\uh - 2n\eta) \right]^{2N} \non \\
&  & + \tilde A^{(m_1)}(u)\, \tilde \psi(u)\
\frac{\sinh u}{\sinh (u-2(2n-1)\eta)}
\frac{\cosh(u-2(n-1)\eta)}{\cosh(u-2n\eta)} 
\left[ 2 \sinh (\uh) \cosh(\uh - 2(n-1)\eta)\right]^{2N} \non \\
&  & + \left\{
\sum_{l=1}^{n-1} \left[ z_l(u)\, \psi(u)\, B_l^{(m_l \,, m_{l+1})}(u)
+ \tilde z_l(u)\, \tilde \psi(u)\, \tilde B_l^{(m_l \,, m_{l+1})}(u) \right] \right\} 
\left[2 \sinh(\uh) \cosh(\uh -2n\eta)\right]^{2N}\,,\non\\
\label{LambdaA2odd}
\end{IEEEeqnarray}
where
\begin{align}
A^{(m_1)}(u) &= \frac{Q_{1}(u+2\eta)}{Q_{1}(u-2\eta)}\,, \non\\
B_l^{(m_l \,, m_{l+1})}(u) &= \frac{Q_{l}(u-2(l+2)\eta)\, 
Q_{l+1}(u-2(l-1)\eta)}{Q_{l}(u-2l\eta)\, Q_{l+1}(u-2(l+1)\eta)} \,,\qquad l = 
1\,, \ldots\,, n-2 \,, \non \\
B_{n-1}^{(m_{n-1} \,, m_{n})}(u) &= \frac{Q_{n-1}(u-2(n+1)\eta)\, 
Q_{n}(u-2(n-2)\eta)}{Q_{n-1}(u-2(n-1)\eta)\, Q_{n}(u-2n\eta)}\,,
\end{align}
with
\begin{align}
Q_{l}(u) &= \prod_{j=1}^{m_{l}}\sinh(\tfrac{1}{2}(u-u_{j}^{(l)}))\, 
\sinh(\tfrac{1}{2}(u+u_{j}^{(l)})) \,, \qquad l = 1, \ldots, n-1 \,, \non\\
Q_{n}(u) &= \prod_{j=1}^{m_{n}}\sinh(u-u_{j}^{(n)})\, \sinh(u+u_{j}^{(n)}) \,, 
\end{align}
and
\be
z_l(u) =  \frac{\sinh(u)}{\sinh(u-2l\eta)}
\frac{\sinh(u-4n\eta)}{\sinh(u-2(l+1)\eta)}
\frac{\cosh(u-2(n+1)\eta)}{\cosh(u-2n\eta)} \,, 
\ee
\be
\psi(u) = \left\{\begin{array}{cl}
1 & \mbox{ for } A_{2n-1}^{(2)} - \text{I}\\ 
- \frac{\cosh(u-2(n-1)\eta)}{\cosh(u-2(n+1)\eta)} & \mbox{ for } 
A_{2n-1}^{(2)} - \text{II}
\end{array} \right. \,.
\label{psiA2odd}
\ee 
The corresponding quantities with tildes are obtained from crossing
\begin{align}
\tilde A^{(m_{1})}(u) &= A^{(m_{1})}(-u-\rho) \,, \qquad \tilde B_{l}^{(m_l \,, m_{l+1})}(u) = 
B_{l}^{(m_l \,, m_{l+1})}(-u-\rho) \,, \non\\
\tilde z_l(u) &= z_{l}(-u-\rho)\,,  \qquad
\qquad \tilde \psi(u) = \psi(-u-\rho) \,.
\label{tildeA2odd}
\end{align}
The results for cases I and II differ only by the function $\psi(u)$ 
(\ref{psiA2odd}). For $\psi(u)=1$, the above expression 
reduces to the result in \cite{Artz:1995bm}.

The eigenvalues of both Hamiltonians ${\cal H}^{(A_{2n-1}^{(2)} - 
\text{I})}$ (\ref{HamiltonianA2oddI}) and 
${\cal H}^{(A_{2n-1}^{(2)} - \text{II})}$ (\ref{HamiltonianA2oddII}) are given 
by
\begin{align}
E &= -\sum_{k=1}^{m_{1}}\frac{\sinh(2\eta)}{2 \sinh(\frac{1}{2} u^{(1)}_{k}-\eta) 
\sinh(\frac{1}{2} u^{(1)}_{k}+\eta)} - 
\frac{(N-1)\cosh(2(n+1)\eta)}{2\sinh(2\eta)\, \cosh(2n\eta)} \qquad & \mbox{  for  } n>1 \,, \non \\
&= -\sum_{k=1}^{m_{1}}\frac{\sinh(4\eta)}{\sinh(u^{(1)}_{k}-2\eta) 
\sinh(u^{(1)}_{k}+2\eta)} - 
\frac{(N-1)\cosh(4\eta)}{\sinh(4\eta)}   & \mbox{  for  } n=1  \,,
\label{BAenergyA2odd}
\end{align}
as follows from (\ref{HA2oddIt})-(\ref{HA2Itc1c2}), 
(\ref{HA2oddIIt})-(\ref{HA2IItc1c2}) and 
(\ref{LambdaA2odd})-(\ref{psiA2odd}).

\subsection{Bethe equations}

The conditions for the cancellation of the poles of $\Lambda^{(m_1 \,, 
\cdots \,, m_n)}(u)$ (\ref{LambdaA2odd})
at $u=u_{k}^{(l)}+ 2 l \eta\,, \ l= 1, \ldots, n$, which are the so-called Bethe equations, are 
given for $n>2$ by
\begin{align}
f_{2}^{2N}(u^{(1)}_{k}) &= \prod_{j=1,\, j\ne k}^{m_{1}}
f_{4}(u^{(1)}_{k}-u^{(1)}_{j})\, f_{4}(u^{(1)}_{k}+u^{(1)}_{j})
\prod_{j=1}^{m_{2}}
f_{-2}(u^{(1)}_{k}-u^{(2)}_{j})\, f_{-2}(u^{(1)}_{k}+u^{(2)}_{j}) \,, 
\non \\
&  \qquad\qquad\qquad\qquad k = 1, \ldots, m_{1}\,, \non \\
1 &= \prod_{j=1}^{m_{l-1}}
f_{-2}(u^{(l)}_{k}-u^{(l-1)}_{j})\, f_{-2}(u^{(l)}_{k}+u^{(l-1)}_{j})
\prod_{j=1,\, j\ne k}^{m_{l}}
f_{4}(u^{(l)}_{k}-u^{(l)}_{j})\, f_{4}(u^{(l)}_{k}+u^{(l)}_{j}) \non \\
&  \times
\prod_{j=1}^{m_{l+1}}
f_{-2}(u^{(l)}_{k}-u^{(l+1)}_{j})\, f_{-2}(u^{(l)}_{k}+u^{(l+1)}_{j})
\,, \quad k=1, \ldots, m_{l}\,, \quad l = 2, \ldots, n-2\,, \non \\
1 &= \prod_{j=1}^{m_{n-2}}
f_{-2}(u^{(n-1)}_{k}-u^{(n-2)}_{j})\, f_{-2}(u^{(n-1)}_{k}+u^{(n-2)}_{j})
\prod_{j=1,\, j\ne k}^{m_{n-1}}
f_{4}(u^{(n-1)}_{k}-u^{(n-1)}_{j})\, f_{4}(u^{(n-1)}_{k}+u^{(n-1)}_{j}) \non \\
&  \times
\prod_{j=1}^{m_{n}}
e_{-2}(u^{(n-1)}_{k}-u^{(n)}_{j})
e_{-2}(u^{(n-1)}_{k}+u^{(n)}_{j})\,,  \qquad k=1, \ldots, m_{n-1}\,,  \non \\
\chi(u^{(n)}_{k}) &= \prod_{j=1}^{m_{n-1}}
e_{-2}(u^{(n)}_{k}-u^{(n-1)}_{j})
e_{-2}(u^{(n)}_{k}+u^{(n-1)}_{j}) \prod_{j=1,\, j\ne k}^{m_{n}}
e_{4}(u^{(n)}_{k}-u^{(n)}_{j})
e_{4}(u^{(n)}_{k}+u^{(n)}_{j})\,,\non \\
&  \qquad\qquad\qquad\qquad k = 1, \ldots, m_{n}\,,
\label{BAEA2odd}
\end{align}
where we use here the compact notation
\be
e_{k}(u) = \frac{\sinh(u + k \eta)}{\sinh(u - k \eta)} \,, \qquad 
f_{k}(u) = \frac{\sinh(\tfrac{1}{2}(u + k \eta))}{\sinh(\tfrac{1}{2}(u - k 
\eta))} \,,
\ee
and
\be
\chi(u) = \left\{ \begin{array}{cr}
1 & \mbox{ for } A_{2n-1}^{(2)} - \text{I}\\ 
\left(\frac{\cosh(u-2\eta)}{\cosh(u+2\eta)}\right)^{2} & \mbox{ for } A_{2n-1}^{(2)} - \text{II}
\end{array}\right. \,.
\ee
For $n=1$, the Bethe equations are given by
\be 
e_{2}^{2N}(u^{(1)}_{k})\, \chi(u^{(1)}_{k})  = \prod_{j=1,\, j\ne k}^{m_{1}}
e_{4}(u^{(1)}_{k}-u^{(1)}_{j})\,
e_{4}(u^{(1)}_{k}+u^{(1)}_{j})\,, \quad k = 1, \ldots, m_{1} \,;
\label{BAEA2oddn1}
\ee
while for $n=2$, the Bethe equations are given by
\begin{align}
f_{2}^{2N}(u^{(1)}_{k}) &= \prod_{j=1,\, j\ne k}^{m_{1}}
f_{4}(u^{(1)}_{k}-u^{(1)}_{j})\, f_{4}(u^{(1)}_{k}+u^{(1)}_{j}) 
\prod_{j=1}^{m_{2}}e_{-2}(u^{(1)}_{k}-u^{(2)}_{j})\, 
e_{-2}(u^{(1)}_{k}+u^{(2)}_{j}) \,, \non \\
& \qquad\qquad\qquad\qquad k = 1, \ldots, m_{1} \,, \non \\
 \chi(u^{(2)}_{k})  &= \prod_{j=1}^{m_{1}}
e_{-2}(u^{(2)}_{k}-u^{(1)}_{j})
e_{-2}(u^{(2)}_{k}+u^{(1)}_{j}) \prod_{j=1,\, j\ne k}^{m_{2}}
e_{4}(u^{(2)}_{k}-u^{(2)}_{j})
e_{4}(u^{(2)}_{k}+u^{(2)}_{j})\,,  \non \\
& \qquad\qquad\qquad\qquad k = 1, \ldots, m_{2} \,.
\label{BAEA2oddn2}
\end{align}

\subsection{Dynkin labels of the Bethe states}

The eigenvalues of the Cartan generators are given by \cite{Artz:1995bm, Ahmed:2017mqq, Reshetikhin:1987}
\begin{align}
    h_{1} & = N - m_{1}\,, \non \\
    h_{i} & = m_{i-1} - m_{i}\,,  \qquad i =2, \ldots, n-1 \,, \non \\
    h_{n} & = m_{n-1} - 2m_{n}\,. 
\end{align}
Using the relation of the $C_{n}$ and $D_{n}$ Dynkin labels $[a_{1},
\ldots, a_{n}]$ to the eigenvalues of the Cartan generators \cite{Ahmed:2017mqq}
\begin{align}
a_{i} &= h_{i} - h_{i+1} \,, \qquad i = 1, 2, \ldots, n-1 \,, \non \\ 
a_{n} &= \left\{ \begin{array}{cc}
h_{n} & \mbox{ for } U_{q}(C_{n})\\
h_{n-1} + h_{n} & \mbox{ for } U_{q}(D_{n})
\end{array} \right. \,,
\label{DynkinCartnrelationCnDn}
\end{align}
we obtain a formula for the Dynkin labels in terms of the 
cardinalities $(m_{1},\ldots, m_{n})$ of the Bethe roots
\begin{align}
a_{1} &= N - 2m_{1} + m_{2} \,, \non \\
a_{i} &= m_{i-1} - 2m_{i} + m_{i+1} \,, \qquad i = 2, \ldots, n-2 
\,, \non \\
a_{n-1} &= m_{n-2} -2 m_{n-1}+ 2 m_{n} \,, \non \\
a_{n} &=  \left\{ \begin{array}{cc}
m_{n-1} - 2 m_{n}&  \mbox{ for } U_{q}(C_{n})\\
m_{n-2} - 2 m_{n} & \mbox{ for } U_{q}(D_{n})
\end{array} \right. \,.
\label{DynkinBArltnCnDn}
\end{align}
The above formulas are for $n>2$; for smaller values of $n$, we obtain
\begin{align}
  \mbox{ for } U_{q}(C_{1}) &: a_{1}=N-2m_{1} \,, \non \\
  \mbox{ for } U_{q}(C_{2}) &: a_{1}=N-2m_{1}+ 2m_{2} \,, \quad a_{2} = 
  m_{1}-2m_{2} \,, \non\\
  \mbox{ for } U_{q}(D_{2}) &: a_{1}=N-2m_{1}+ 2m_{2} \,, \quad a_{2} = 
  N-2m_{2} \,.
\end{align}

\subsection{Numerical check of 
completeness}\label{sec:completenessA2odd}

We present solutions ($\{u^{(1)}_{1}, \ldots, 
u^{(1)}_{m_{1}}\}, \ldots, \{u^{(n)}_{1}, \ldots, 
u^{(n)}_{m_{n}}\} $)
of the $A^{(2)}_{2n-1}$
Bethe equations (\ref{BAEA2odd})-(\ref{BAEA2oddn2}) for small 
values of $n$ and $N$ and a generic value of $\eta$ (namely, 
$\eta = 
-i/10$) in Tables \ref{table:C1N2}-\ref{table:C3N3} for case I  (\ref{KA2oddsetI}), 
and in Tables \ref{table:D1N2}-\ref{table:D3N3} for case II 
(\ref{KA2oddsetII}).\footnote{The Bethe equations are invariant under the reflections $u^{(l)}_{k} \mapsto 
-u^{(l)}_{k}$, as well as under the shifts $u^{(l)}_{k} \mapsto 
u^{(l)}_{k} + 2\pi i $ (except for $l=n$, in which case the shift symmetry is  
$u^{(n)}_{k} \mapsto u^{(n)}_{k} + \pi i$).
The Bethe roots can therefore be restricted to the domain
$\Im m (u^{(l)}_{k}) \in [0, 2\pi)$ (except for $l=n$, in 
which case $\Im m (u^{(l)}_{k}) \in [0, \pi)$),
and $\Re e  (u^{(l)}_{k}) \ge 0$.}
Each table also displays the cardinalities $(m_{1}, \ldots, m_{n})$ 
of the Bethe roots and the degeneracy (``deg'') of the corresponding 
eigenvalue of the Hamiltonians ${\cal H}^{(A_{2n-1}^{(2)} - 
\text{I})}$ and ${\cal H}^{(A_{2n-1}^{(2)} - 
\text{II})}$
(or, equivalently, of the transfer matrix $t(u)$ at some generic 
value of $u$) obtained by direct diagonalization. For cases with quantum group 
symmetry, the tables also display
the corresponding Dynkin label $[a_{1}, \ldots, a_{n}]$ obtained using the formula 
(\ref{DynkinBArltnCnDn}), and the 
multiplicity (``mult'') i.e., the number of solutions of the Bethe equations 
with the given cardinality of Bethe roots.

For case I (Tables \ref{table:C1N2}-\ref{table:C3N3}), 
the degeneracies exactly coincide with the dimensions of the 
representations corresponding to the Dynkin labels.\footnote{The dimensions 
corresponding to the Dynkin labels can be read off from the tensor 
product decompositions (\ref{decompC1})-(\ref{decompC3}), 
or more generally can be obtained from e.g. \cite{Feger:2012bs}.}
Moreover, the degeneracies and multiplicities predicted by the $U_{q}(C_{n})$ 
symmetry (\ref{decompC1})-(\ref{decompC3}) are completely accounted for by the 
Bethe ansatz solutions. 

Case II is less straightforward. Since the $U_{q}(D_{n})$ symmetry first appears for 
$n=2$, we present for $n=1$ (Tables \ref{table:D1N2}-\ref{table:D1N3}) 
only the Bethe roots and the degeneracies of the corresponding 
eigenvalues, which accounts for all $2^{N}$ eigenvalues. For 
$n=2$ and $n=3$ (Tables \ref{table:D2N2}-\ref{table:D3N3}), certain degeneracies 
are higher than expected from the $U_{q}(D_{n})$ symmetry, and are 
indicated with a star $(*)$. We expect that these higher degeneracies 
are due to an additional $Z_{2}$ symmetry relating representations to 
their conjugates, see Section \ref{sec:degmultDn}.

The eigenvalues of the Hamiltonians ${\cal H}^{(A_{2n-1}^{(2)} -
\text{I})}$ and ${\cal H}^{(A_{2n-1}^{(2)} - \text{II})}$, as well as
the eigenvalues of the transfer matrix $t(u)$ (\ref{transfer}) for the
two cases (\ref{KA2oddsetI})-(\ref{KA2oddsetII}) at some generic value
of $u$, are not displayed in the tables in order to minimize their
size.  Nevertheless, we have computed these eigenvalues both directly
and from the displayed solutions of the Bethe equations using
(\ref{BAenergyA2odd}) and (\ref{LambdaA2odd})-(\ref{tildeA2odd}),
respectively; and we find perfect agreement between the results from
these two approaches.

\section{Bethe ansatz for $D_{n+1}^{(2)}$ - case I}\label{sec:BAD2}

Among the infinite families of anisotropic R-matrices associated with
affine Lie algebras that were found in \cite{Bazhanov:1984gu,
Bazhanov:1986mu, Jimbo:1985ua, Kuniba:1991yd}, the case
$D_{n+1}^{(2)}$ (\ref{RmatD2}) is by far the most complicated.
Moreover, the corresponding K-matrices (\ref{KD2}) are also 
complicated, since they are not diagonal. Therefore, it is not 
surprising that little is known about the eigenvalues of the 
corresponding transfer matrices. (The eigenvalues of the Hamiltonian 
for case I with $n=1$ were determined in \cite{Martins:2000xie}.)

In this section, we propose an expression for the eigenvalues of the 
transfer matrix for $D_{n+1}^{(2)}$ - case I (\ref{QGKMI}). We also propose a 
formula for the Dynkin labels of a Bethe state in terms of the 
cardinalities of the corresponding Bethe roots. Moreover, we check the completeness 
of our Bethe ansatz solution numerically for small values of $n$ and $N$.
(Unfortunately, we have not succeeded to find similarly satisfactory results for case II.)

\subsection{Transfer matrix eigenvalues}
 
For real values of $\eta$, the transfer matrix (\ref{transfer}) for $D_{n+1}^{(2)}$ - case I 
is Hermitian. The commutativity property (\ref{commutativity}) and the fact that the 
transfer matrix also commutes with all of the Cartan generators 
(\ref{transfCartanBn}) imply that the transfer matrix and Cartan 
generators can be simultaneously diagonalized
\begin{align}
t(u)\, |\Lambda^{(m_{1}, \ldots, m_{n})}\rangle &= \Lambda^{(m_{1}, 
\ldots, m_{n})}(u)\,
|\Lambda^{(m_{1}, \ldots, m_{n})}\rangle \,, \non \\
\Delta_{N}(H_{j})\, |\Lambda^{(m_{1}, \ldots, m_{n})}\rangle &= h_{j}\,
|\Lambda^{(m_{1}, \ldots, m_{n})}\rangle \,, \qquad j = 1, \ldots, n 
\,,
\end{align}
where the Bethe states $|\Lambda^{(m_{1}, \ldots, m_{n})}\rangle$ are 
independent of the spectral parameter. We assume that the Bethe 
states are highest-weight states of $U_{q}(B_{n})$
\be
\Delta_{N}(E^{+}_{j})\, |\Lambda^{(m_{1}, \ldots, m_{n})}\rangle =0 \,, 
\qquad j = 1, \ldots, n 
\,.
\label{highestweightD2}
\ee

We now determine the eigenvalues $\Lambda^{(m_{1}, \ldots, m_{n})}(u)$ 
of the transfer matrix using an analytical Bethe ansatz 
approach \cite{Mezincescu:1991ag, Artz:1994qy, Artz:1995bm, Vichirko1983, 
Reshetikhin:1983vw, Reshetikhin:1987, Wang2015}. Recalling the 
result for the periodic chain \cite{Reshetikhin:1987}, the crossing 
symmetry (\ref{crossingtransfer}), and assuming 
the ``doubling hypothesis'' \cite{Artz:1994qy, Artz:1995bm}, we arrive at 
the following expression for the transfer matrix eigenvalues\footnote{We note 
the following typographical errors in \cite{Reshetikhin:1987}: in 
the second line of (9), $\lambda \eta \mapsto \alpha \lambda \eta$; 
and in the second equation in (23) (i.e., for $B_{n-2}(u)$), $Q_{n-2} 
\mapsto Q_{n-2}^{2}$.}
\begin{align}
\Lambda^{(m_{1}, \ldots, m_{n})}(u) &= a(u)\, A(u)\, [4 \sinh(u-2\eta) 
\sinh(u-2 n \eta) ]^{2N} \non\\
&  + \left\{\sum_{l=1}^{n} \left[ b_{l}(u)\, B_{l}(u) +
\tilde b_{l}(u)\, \tilde B_{l}(u) \right] \right\} [4 \sinh(u) 
\sinh(u-2 n \eta) ]^{2N}\non\\
&  + \tilde a(u)\, \tilde A(u)\, [4 \sinh(u) 
\sinh(u-2 (n-1) \eta) ]^{2N} \,,
\label{Lambda}
\end{align}
where
\begin{align}
A(u) &= \frac{Q_{1}(u+\eta)\, Q_{1}(u+\eta+i \pi)}{Q_{1}(u-\eta)\, 
Q_{1}(u-\eta+i \pi)} \,, \non \\
B_{l}(u) &= \frac{Q_{l}(u-(l+2)\eta)\, Q_{l}(u-(l+2)\eta+i 
\pi)}{Q_{l}(u-l \eta)\, 
Q_{l}(u-l \eta+i \pi)}\non \\
& \times \frac{Q_{l+1}(u-(l-1)\eta)\, Q_{l+1}(u-(l-1)\eta+i 
\pi)}{Q_{l+1}(u-(l+1) \eta)\, 
Q_{l+1}(u-(l+1) \eta+i \pi)}
\,, \qquad l = 1, \ldots, n-2 \,, \non \\
B_{n-1}(u) &=  \frac{Q_{n-1}(u-(n+1)\eta)\, Q_{n-1}(u-(n+1)\eta+i \pi)\,Q_{n}(u-(n-2)\eta) \, Q_{n}(u-(n-2)\eta+i \pi)}
{Q_{n-1}(u-(n-1)\eta)\, Q_{n-1}(u-(n-1)\eta+i \pi)\, Q_{n}(u-n\eta) \, Q_{n}(u-n\eta+i \pi)} \,, \non \\
B_{n}(u) &=  \frac{Q_{n}(u-(n+2)\eta)\, Q_{n}(u-(n-2)\eta+i \pi)}
{Q_{n}(u-n\eta)\, Q_{n}(u-n\eta+i \pi)} \,,
\label{ABfuncs}
\end{align}
with
\be
Q_{l}(u) = \prod_{j=1}^{m_{l}}\sinh(\tfrac{1}{2}(u-u_{j}^{(l)}))\, 
\sinh(\tfrac{1}{2}(u+u_{j}^{(l)})) \,.
\label{QD2}
\ee
Moreover,
\begin{align}
b_{1}(u) &= a(u)\, \frac{\sinh(2u)}{\sinh(2u-4\eta)} \,, \non \\
b_{l}(u) &= b_{l-1}(u)\, \frac{\sinh(2u-2(l-1)\eta)}
{\sinh(2u-2(l+1)\eta)} \,, \qquad l = 2, \ldots, n-1 \,, \non \\
b_{n}(u) &= b_{n-1}(u)\, 
\frac{\sinh(u-(n-1)\eta)}{\sinh(u-(n+1)\eta)}  \,.
\label{bsD2}
\end{align}
For $n=1$, only the final equation in (\ref{bsD2}) applies, with $b_{0}(u) \equiv a(u)$.
The corresponding quantities with tildes are obtained from crossing
\begin{align}
\tilde A(u) &= A(-u-\rho) \,, \qquad \tilde a(u) = a(-u-\rho) \non \\
\tilde B_{l}(u) &= B_{l}(-u-\rho) \,, \qquad \tilde b_{l}(u) = 
b_{l}(-u-\rho)\,,  \qquad l = 1, \ldots, n \,.
\label{tildeD2}
\end{align}

\subsection{Determining $a(u)$}\label{sec:afunc}

There remains to determine the function $a(u)$. For small values of 
$n$, this function can be easily found by directly computing the 
eigenvalues of the transfer matrix corresponding to the reference state 
$\scriptscriptstyle{\left(\begin{array}{c}
1\\
0\\
\vdots\\
0
\end{array}\right)}^{\otimes N}_{2n+2}$ for $N=0,1, \ldots$, and comparing 
these results with the expression from (\ref{Lambda})
\begin{align}
\Lambda^{(0, \ldots, 0)}(u) &= a(u)\, [4 \sinh(u-2\eta) 
\sinh(u-2 n \eta) ]^{2N} \non\\
&  + \left\{\sum_{l=1}^{n} \left[ b_{l}(u) +
\tilde b_{l}(u) \right] \right\} [4 \sinh(u) 
\sinh(u-2 n \eta) ]^{2N}\non\\
&  + \tilde a(u)\, [4 \sinh(u) 
\sinh(u-2 (n-1) \eta) ]^{2N} \,.
\end{align}
In this way, we obtain
\be
a(u) =\left\{\begin{array}{cc}
\frac{4 e^{6\eta}\cosh(u-\eta) \cosh(u) \sinh(2(u-2 \eta)) \sinh(u-2\eta)}
{\sinh(2(u-\eta)) \sinh(u-\eta)} & 
\mbox{ for } n=1 \\
\frac{4 e^{12 \eta}\cosh(u-2\eta) \cosh(u-\eta) \sinh(2(u-4 \eta)) \sinh(u-3\eta)}
{\sinh(2(u-\eta)) \sinh(u-2\eta)} & 
\mbox{ for } n=2
\end{array} \right. \,.
\label{afuncn12}
\ee

For general values of $n$, the function $a(u)$ can be 
determined with the help of the functional equation obeyed by the {\it inhomogeneous} transfer 
matrix \cite{Wang2015}. We therefore introduce inhomogeneities $\{\theta_{j}\}$, 
so that the corresponding transfer matrix is given by
\be
t(u; \{\theta_{j}\}) = \tr_a K^{+}_a(u)\, T_a(u; \{\theta_{j}\})\,  
K^{-}_{a}(u)\, \hat T_a(u; \{\theta_{j}\}) \,, 
\label{transferinhomo}
\ee
where
\begin{align}
T_a(u; \{\theta_{j}\}) &= R_{aN}(u-\theta_{N})\ R_{a 
N-1}(u-\theta_{N-1})\ \cdots R_{a1}(u-\theta_{1}) \,,  \non \\
\hat T_a(u; \{\theta_{j}\})  &= R_{1a}(u+\theta_{1})\ \cdots R_{N-1 a}(u+\theta_{N-1})\ 
R_{Na}(u+\theta_{N}) \,.
\label{monodromyinhomo}
\end{align}
Using the fusion procedure \cite{Kulish:1981gi, Kulish:1981bi}, as 
generalized to the case with boundaries in \cite{Mezincescu:1991ke}, 
we obtain the fusion formula
\be
t(u; \{\theta_{j}\})\,  t(u+\rho; \{\theta_{j}\}) = 
\frac{1}{\zeta(2u+2\rho)}\left[\tilde t(u;  \{\theta_{j}\}) + 
f_{0}(u)\, f_{1}(u)\, \id^{\otimes 
N}\right]\,,
\ee
where $\tilde t(u; \{\theta_{j}\})$ is a fused transfer matrix. 
Moreover, the scalar functions $f_{i}(u)$ are given by products of quantum 
determinants
\begin{align}
f_{0}(u) &=  \delta(T(u; 
\{\theta_{j}\})\,\delta(\hat T(u; \{\theta_{j}\}) \non \\
&= \prod_{k=1}^{N} \zeta(u-\theta_{k}+\rho)\, 
\zeta(u+\theta_{k}+\rho)\,,
\end{align}
and
\begin{align}
f_{1}(u) &= \delta(K^{+}(u))\, \delta(K^{-}(u)) \non \\
&= 2^{10} e^{12 n \eta} \cosh^{2}(u-3 n \eta)  \cosh^{2}(u- n \eta) 
\cosh(u-(n+1)\eta) \cosh(u-(3n-1)\eta) \non\\
& \times \sinh(2u) \sinh(u-(n-1)\eta) 
\sinh(2(u-4 n \eta)) \sinh(u-(3n+1) \eta) \,.
\end{align}
Using the fact that the fused transfer matrix vanishes at 
$u=\theta_{i}-\rho$
\be
\tilde t(u;  \{\theta_{j}\})\Big\vert_{u=\theta_{i}-\rho} = 0 \,,
\ee
we obtain a functional relation for the fundamental transfer matrix
\be
t(\theta_{i}-\rho; \{\theta_{j}\})\,  t(\theta_{i}; \{\theta_{j}\}) = 
\frac{f_{0}(\theta_{i}-\rho)\, f_{1}(\theta_{i}-\rho)}{\zeta(2\theta_{i})} \, \id^{\otimes 
N}\,,
\ee
which implies a corresponding result for the eigenvalues
\be
\Lambda^{(m_{1}, \ldots, m_{n})}(-\theta_{i}; \{\theta_{j}\})\,  
\Lambda^{(m_{1}, \ldots, m_{n})}(\theta_{i}; \{\theta_{j}\}) = 
\frac{f_{0}(\theta_{i}-\rho)\, f_{1}(\theta_{i}-\rho)}{\zeta(2\theta_{i})}\,,
\label{funcrltn1}
\ee
where we have again used the crossing symmetry 
(\ref{crossingtransfer}). The expression for the eigenvalues
of the transfer matrix in the presence of inhomogeneities
$\Lambda^{(m_{1}, \ldots, m_{n})}(u; \{\theta_{j}\})$
is the same as 
(\ref{Lambda}), except with the following replacements
\begin{align}
[4 \sinh(u-2\eta)\sinh(u-2 n \eta) ]^{2N} &\mapsto  
\prod_{k=1}^{N} 
[4 \sinh(u-\theta_{k}-2\eta) \sinh(u-\theta_{k}-2 n \eta) ] \non \\
& \qquad \times
[4 \sinh(u+\theta_{k}-2\eta) \sinh(u+\theta_{k}-2 n \eta) ]
\non \\
[4 \sinh(u) \sinh(u-2 n \eta) ]^{2N} &\mapsto 
\prod_{k=1}^{N}
[4 \sinh(u-\theta_{k}) \sinh(u-\theta_{k}-2 n \eta) ] \non \\
& \qquad \times
[4 \sinh(u+\theta_{k}) \sinh(u+\theta_{k}-2 n \eta) ] \non \\
[4 \sinh(u) \sinh(u-2 (n-1) \eta) ]^{2N} 
&\mapsto  \prod_{k=1}^{N}
[4 \sinh(u-\theta_{k}) \sinh(u-\theta_{k}-2 (n-1) \eta) ] \non \\
& \qquad \times
[4 \sinh(u+\theta_{k}) \sinh(u+\theta_{k}-2 (n-1) \eta) ]  \,.
\end{align}
Notice that the latter two expressions vanish for $u=\pm \theta_{i}$.
The functional relation (\ref{funcrltn1}) therefore reduces to a 
functional relation for the unknown function $a(u)$
\be
a(u)\, a(-u) = \frac{f_{1}(u-\rho)}{\zeta(2u)} \,.
\label{aafuncrltn}
\ee 
A solution of the functional relation (\ref{aafuncrltn}), which 
agrees with the results for $n=1$ 
and $n=2$ (\ref{afuncn12}), is given by
\be
a(u) = \frac{4 e^{6 n \eta}\cosh(u-n \eta) \cosh(u-(n-1)\eta) 
\sinh(2(u-2n \eta)) \sinh(u-(n+1)\eta)}{\sinh(2(u-\eta)) \sinh(u-n 
\eta)} \,.
\label{afunc}
\ee

In summary, the eigenvalues of the transfer matrix for 
$D_{n+1}^{(2)}$ - case I are given by (\ref{Lambda})-(\ref{tildeD2}) 
and (\ref{afunc}). The eigenvalues of the Hamiltonian ${\cal H}^{(D_{n+1}^{(2)} - 
\text{I})}$ (\ref{HamiltonianD2I}) are therefore given by 
\be
E =-\sum_{k=1}^{m_{1}}\frac{\sinh(2\eta)}{\sinh(u^{(1)}_{k}-\eta) 
\sinh(u^{(1)}_{k}+\eta)} - 
\frac{(N-1)\sinh(2(n+1)\eta)}{\sinh(2\eta)\, \sinh(2n\eta)} \,,
\label{BAenergyD2I}
\ee
as follows from (\ref{HD2It})-(\ref{HD2Itc1c2}), 
(\ref{Lambda})-(\ref{tildeD2}) and (\ref{afunc}).

\subsection{Bethe equations}

The corresponding Bethe equations can be obtained by demanding the cancellation of 
the poles in $\Lambda^{(m_{1}, \ldots, m_{n})}(u)$ (\ref{Lambda}) at 
$u=u_{k}^{(l)}+ l \eta\,, \ l= 1, \ldots, n$. In terms of the 
notation
\be
e_{k}(u) = \frac{\sinh(u + k \eta)}{\sinh(u - k \eta)} \,, \qquad 
f_{k}(u) = \frac{\sinh(\tfrac{1}{2}(u + k \eta))}{\sinh(\tfrac{1}{2}(u - k 
\eta))} \,,
\label{defsef}
\ee
the Bethe equations for $n>1$ are given by
\begin{align}
e_{1}^{2N}(u_{k}^{(1)}) &= \prod_{j=1, j\ne k}^{m_{1}} 
e_{2}(u_{k}^{(1)}-u_{j}^{(1)})\, e_{2}(u_{k}^{(1)}+u_{j}^{(1)})
\prod_{j=1}^{m_{2}} 
e_{-1}(u_{k}^{(1)}-u_{j}^{(2)})\, e_{-1}(u_{k}^{(1)}+u_{j}^{(2)}) \,, 
\non \\
& \qquad\qquad\qquad\qquad\qquad k = 1, \ldots, m_{1}\,, \non\\ 
1  &= \prod_{j=1, j\ne k}^{m_{l}} 
e_{2}(u_{k}^{(l)}-u_{j}^{(l)})\, e_{2}(u_{k}^{(l)}+u_{j}^{(l)})
\prod_{j=1}^{m_{l-1}} 
e_{-1}(u_{k}^{(l)}-u_{j}^{(l-1)})\, e_{-1}(u_{k}^{(l)}+u_{j}^{(l-1)}) 
\non \\
& \qquad \times \prod_{j=1}^{m_{l+1}} 
e_{-1}(u_{k}^{(l)}-u_{j}^{(l+1)})\, e_{-1}(u_{k}^{(l)}+u_{j}^{(l+1)}) 
\,, \non\\ 
& \qquad\qquad\qquad\qquad\qquad  k = 1, \ldots, m_{l}\,, \qquad l=2, \ldots, n-1\,, \non \\
1  &= \prod_{j=1, j\ne k}^{m_{n}} 
f_{2}(u_{k}^{(n)}-u_{j}^{(n)})\, f_{2}(u_{k}^{(n)}+u_{j}^{(n)})
\prod_{j=1}^{m_{n-1}} 
e_{-1}(u_{k}^{(n)}-u_{j}^{(n-1)})\, e_{-1}(u_{k}^{(n)}+u_{j}^{(n-1)}) \,, \non\\ 
& \qquad\qquad\qquad\qquad\qquad k = 1, \ldots, m_{n} \,.
\label{BAED2}
\end{align}
For $n=1$, the Bethe equations are given by\footnote{The Bethe 
equations for the case $n=1$ have been proposed 
by Martins and Guan on the basis of a coordinate Bethe 
ansatz analysis; their result (see Eq. (50) in \cite{Martins:2000xie}) is missing the restriction ($j\ne k$) on the product, but
is otherwise equivalent to (\ref{BAEn1}).} 
\be
e_{1}^{2N}(u_{k}^{(1)}) = \prod_{j=1, j\ne k}^{m_{1}} 
f_{2}(u_{k}^{(1)}-u_{j}^{(1)})\, f_{2}(u_{k}^{(1)}+u_{j}^{(1)}) \,, 
\qquad k = 1, \ldots, m_{1}\,.
\label{BAEn1}
\ee
Note that the Bethe equations are exactly ``doubled'' with respect to those for the 
periodic chain \cite{Reshetikhin:1987}. Indeed, the functions 
$b_{l}(u)$ in (\ref{Lambda}) were ``reverse engineered'' in 
(\ref{bsD2}) to obtain this result.

\subsection{Dynkin labels of the Bethe states}

The eigenvalues of the Cartan generators are given by
\begin{align}
    h_{1} & = N - m_{1}\,, \non \\
    h_{i} & = m_{i-1} - m_{i}\,,  \qquad i =2, \ldots, n \,.
\end{align}
Using the relation of the $B_{n}$ Dynkin label $[a_{1},
\ldots, a_{n}]$ to the eigenvalues of the Cartan generators \cite{Ahmed:2017mqq}
\begin{align}
a_{i} &= h_{i} - h_{i+1} \,, \qquad i = 1, 2, \ldots, n-1 \,, \non \\ 
a_{n} &= 2 h_{n} \,,
\label{DynkinCartnrelation}
\end{align}
we obtain a formula for the Dynkin label in terms of the 
cardinalities $(m_{1},\ldots, m_{n})$ of the Bethe roots
\begin{align}
a_{1} &= N - 2m_{1} + m_{2} \,, \non \\
a_{i} &= m_{i-1} - 2m_{i} + m_{i+1} \,, \qquad i = 2, \ldots, n-1 
\,, \non \\
a_{n} &= 2 (m_{n-1} - m_{n}) \,.
\label{DynkinBArltn}
\end{align}
The above formula is for $n>1$; for $n=1$, we obtain $a_{1}=2(N-m_{1})$.

\subsection{Numerical check of completeness}\label{sec:completenessD2}

We present solutions ($\{u^{(1)}_{1}, \ldots, 
u^{(1)}_{m_{1}}\}, \ldots, \{u^{(n)}_{1}, \ldots, 
u^{(n)}_{m_{n}}\} $)
of the Bethe equations for $D_{n+1}^{(2)}$ - case I
(\ref{BAED2})-(\ref{BAEn1}) for small 
values of $n$ and $N$ and a generic value of $\eta$ (namely, $\eta = 
-i/10$) in Tables \ref{table:B1N2}-\ref{table:B2N3}.\footnote{
The Bethe equations are invariant under the reflections $u^{(l)}_{k} \mapsto 
-u^{(l)}_{k}$, as well as under the shifts $u^{(l)}_{k} \mapsto u^{(l)}_{k} + \pi i$
(except for $l=n\,, \ m_{n}>1$, in which case the shift symmetry is only 
$u^{(n)}_{k} \mapsto u^{(n)}_{k} + 2\pi i$).
% \begin{align}
%     & n=1\,, \ m_{1}=1: & u^{(1)}_{1} \mapsto u^{(1)}_{1} + \pi i \,, \non\\
%     & n=1\,, \ m_{1}>1: & u^{(1)}_{k} \mapsto u^{(1)}_{k} + 2\pi i \,, \non\\
%     & n>1\,, \ l<n:     & u^{(l)}_{k} \mapsto u^{(l)}_{k} + \pi i \,, \non\\
%     & n>1\,, \ l=n\,, \ m_{n}=1: & u^{(n)}_{1} \mapsto u^{(n)}_{1} + \pi i \,, \non\\
%     & n>1\,, \ l=n\,, \ m_{n}>1: & u^{(n)}_{k} \mapsto u^{(n)}_{k} + 2\pi  
%     i \,.
% \end{align}    
The Bethe roots can therefore be restricted to the domain
$\Im m (u^{(l)}_{k}) \in [0, \pi)$ (except for $l=n\,, \ m_{n}>1$, in 
which case $\Im m (u^{(l)}_{k}) \in [0, 2\pi)$),
and $\Re e  (u^{(l)}_{k}) \ge 0$.} Each table also displays the cardinalities $(m_{1}, \ldots, m_{n})$ 
of the Bethe roots, the corresponding Dynkin label $[a_{1}, \ldots, a_{n}]$ obtained using the formula 
(\ref{DynkinBArltn}), the degeneracy (``deg'') of the corresponding 
eigenvalue of the Hamiltonian ${\cal H}^{(D_{n+1}^{(2)} - \text{I})}$
(or, equivalently, of the transfer matrix $t(u)$ at some generic 
value of $u$) obtained by direct diagonalization, and the 
multiplicity (``mult'') i.e., the number of solutions of the Bethe equations 
with the given cardinality of Bethe roots.

For $n=2$ (Tables \ref{table:B2N2}-\ref{table:B2N3}), the 
degeneracies almost exactly coincide with the dimensions of the
representations corresponding to the Dynkin labels. The one exception 
is indicated with a star (*). The degeneracies and multiplicities
agree almost exactly with those predicted by the $U_{q}(B_{2})$ symmetry 
(\ref{B2degen}).

However, for $n=1$ (Tables \ref{table:B1N2}-\ref{table:B1N3}), many
degeneracies are higher than expected from the
$U_{q}(B_{1})$ symmetry (\ref{B1degen}); these degeneracies are again indicated with 
a star. As already remarked in Section \ref{sec:degmult}, it would be 
interesting to find a symmetry that can account for these higher degeneracies.

The eigenvalues of the Hamiltonian ${\cal H}^{(D_{n+1}^{(2)} -
\text{I})}$ and of the transfer matrix at some generic value of $u$
computed using (\ref{BAenergyD2I}) and (\ref{Lambda})-(\ref{tildeD2})
with the Bethe roots in the tables agree with the eigenvalues 
obtained by direct diagonalization. 

\section{Conclusions}\label{sec:conclusion}

We have identified three infinite families of integrable
open spin chains with quantum group symmetry, corresponding to the following 
three $K^{-}$-matrices: $A_{2n-1}^{(2)}$ - case II
(\ref{KA2oddsetII}), $D_{n+1}^{(2)}$ - case I (\ref{QGKMI}), and
$D_{n+1}^{(2)}$ - case II (\ref{QGKMII}).  We have shown that the
Hamiltonian for $A_{2n-1}^{(2)}$ - case II has the symmetry
$U_{q}(D_{n})$, while both $D_{n+1}^{(2)}$ Hamiltonians have the
symmetry $U_{q}(B_{n})$.  We have proposed Bethe ansatz solutions for
$A_{2n-1}^{(2)}$ - case II and $D_{n+1}^{(2)}$ - case I, whose
completeness we have checked numerically for small values of $n$ and
$N$.  

We have also found formulas for the Dynkin labels in terms of the
cardinalities of the Bethe roots for the latter models, as well as for
$A_{2n-1}^{(2)}$ - case I (\ref{KA2oddsetI}) that has $U_{q}(C_{n})$
symmetry.  For most eigenvalues of the Hamiltonian (or transfer 
matrix) for these models, the degeneracies coincide with the
dimensions of the representations corresponding to the Dynkin labels.
However, we find exceptions, where the degeneracies are higher than 
expected from the quantum group symmetry. (These higher degeneracies 
are designated by a star (*) in Tables \ref{table:C1N2}-\ref{table:B2N3}.) 
It would be interesting to find additional symmetries that can account
for these higher degeneracies.

We did not manage to find a satisfactory Bethe ansatz solution for
$D_{n+1}^{(2)}$ - case II. We expect that for this case the
eigenvalues of the transfer matrix are again given by
(\ref{Lambda})-(\ref{QD2}), but with a different choice for the
functions $a(u)$ and $b_{l}(u)$.  The main remaining difficulty is to determine
$b_{l}(u)$, since the ``doubling hypothesis'' that we used to
determine these functions for case I no longer works.  It is possible
to formulate functional relations for $b_{l}(u)$, whose solutions are
not unique. (For some other examples, see (\ref{b1frdiag}) and 
(\ref{bsum}).)
For the solutions $b_{l}(u)$ that we found, we were
not able to verify completeness even for small values of $n$ and $N$.
It would be interesting to find a set of functions $b_{l}(u)$ for 
case II that does give completeness, as in case I.

An alternative approach for solving $D_{n+1}^{(2)}$ - case II (as well
as other choices of K-matrices) would be algebraic Bethe ansatz, which
would provide not only the eigenvalues but also the eigenvectors of
the transfer matrix.  In principle, this approach is possible, since
the conventional reference state is an eigenstate, despite the fact
that the K-matrices are not diagonal.  Nevertheless, due to the
complexity of both the R-matrix and K-matrices, executing the
algebraic Bethe ansatz for this model would be a nontrivial task.

Additional open problems that were noted for $A_{2n}^{(2)}$ in 
\cite{Ahmed:2017mqq} also apply here, among which are:
proving that the transfer matrix $t(u)$ for $A_{2n-1}^{(2)}$ - case II
and for $D_{n+1}^{(2)}$ - cases I and II also has 
quantum group symmetry; showing that the Bethe states have the 
highest weight property (\ref{highestweightA2odd}), 
(\ref{highestweightD2}); and investigating the case 
that $q$ is a root of unity (non-generic values of $\eta$).

We have completed the program (initiated in \cite{Ahmed:2017mqq}) of 
identifying $K^{-}$-matrices $\ne \id$ associated with the infinite families of 
{\em twisted} affine Lie algebras that can be used to construct
integrable open spin chains with maximal quantum group symmetry. It 
remains to perform a similar search for $K^{-}$-matrices $\ne \id$ associated with
the infinite families of {\em untwisted} affine Lie algebras, in particular $B^{(1)}_{n}$, 
$C^{(1)}_{n}$ and $D^{(1)}_{n}$, which (as expected from the extended 
Dynkin diagrams, as discussed in the Introduction) should give integrable open spin chains with
$U_{q}(D_{n})$, $U_{q}(C_{n})$ and $U_{q}(D_{n})$ symmetry, 
respectively. Of course, this search could be extended even further to $K^{-}$-matrices
associated with the exceptional affine Lie algebras, the affine Lie superalgebras, etc.

In closing, we would like to reiterate that the simplest anisotropic
quantum spin chains are arguably those that are integrable and have quantum
group symmetry.  The main purpose of \cite{Ahmed:2017mqq} and the
present paper was to enlarge the set of such models and their
solutions.  Since simple mathematical models often have nice physical
applications, we expect that these models and their solutions will
find applications to interesting physical problems.
(For recent discussions of applications of periodic $A^{(2)}_{n}$ 
spin chains, see \cite{Vernier:2016xha} and references therein.)

\section*{Acknowledgments}
The work of RN and RP was supported by the S\~ao Paulo Research Foundation (FAPESP) and the University of Miami under the SPRINT
grant \#2016/50023-5. Additional support was provided by a Cooper 
fellowship (RN) and by FAPESP grant \# 2014/00453-8 and FAPESP/CAPES grant 
\# 2017/02987-8 (RP). ALR was supported by the S\~ao Paulo
Research Foundation FAPESP under the process  \# 2017/03072-3 and 
\# 2015/00025-9. RN acknowledges the hospitality at UFSCar and at ICTP-SAIFR. RP thanks the University of Miami for its warm hospitality.

\appendix

\section{R-matrices: explicit formulas}\label{sec:Rmatexplicit}

We present here the explicit R-matrices used in the main text for the convenience of the 
reader.

\subsection{$A_{2n-1}^{(2)}$}\label{sec:RmatA2odd}

We use here the same $A_{2n-1}^{(2)}$ R-matrix that was used in
\cite{Artz:1995bm, Li:2006mv}, which is {\em different} from the $A_{2n-1}^{(2)}$
R-matrix in \cite{Bazhanov:1984gu, Bazhanov:1986mu, Jimbo:1985ua}.  It
can be obtained from the $C_{n}^{(1)}$ R-matrix in \cite{Jimbo:1985ua} by
replacing $\xi=k^{2n+2}$ by $\xi=-k^{2n}$; i.e. by changing $\xi
\mapsto -\xi k^{-2}$.  It is the same as the $A_{2n-1}^{(2)}$ R-matrix
in the appendix of \cite{Kuniba:1991yd} up to some redefinitions of the anisotropy and 
spectral parameters, and an overall factor.

This R-matrix is given by
\begin{align}
R(u)= & c(u)\sum_{\alpha\neq\alpha'}e_{\alpha\alpha}\otimes e_{\alpha\alpha}
+b(u)\sum_{\alpha\neq\beta,\beta'}e_{\alpha\alpha}\otimes e_{\beta\beta}\nonumber\\
& +\left(\,e(u)\sum_{\alpha<\beta,\alpha\neq\beta'}
+\,\bar{e}(u)\sum_{\alpha>\beta,\alpha\neq\beta'}\right)e_{\alpha\beta}\otimes e_{\beta\alpha}
+\sum_{\alpha,\beta}a_{\alpha\beta}(u)e_{\alpha\beta}\otimes e_{\alpha'\beta'}
\label{RA2odd}
\end{align}
\noindent
in which
\begin{align}
& 
c(u)=2\sinh\left(\frac{u}{2}-2\eta\right)\cosh\left(\frac{u}{2}-2n\eta\right)\,,\nonumber\\
& 
b(u)=2\sinh\left(\frac{u}{2}\right)\cosh\left(\frac{u}{2}-2n\eta\right)\,,\nonumber\\
& 
e(u)=-2e^{-\frac{u}{2}}\sinh\left(2\eta\right)\cosh\left(\frac{u}{2}-2n\eta\right)\,,\nonumber\\
& \bar{e}(u)=e^ue(u)\,,
\end{align}
\noindent

and
\begin{equation}
a_{\alpha\beta}(u)=\begin{cases}
2\sinh\left(\frac{u}{2}\right)\cosh\left(\frac{u}{2}-2(n-1)\eta\right)& \alpha=\beta, \alpha\neq\alpha'\\
2\sinh\left(2\eta\right)e^{\mp\frac{u}{2}}
\left[\mp\epsilon_{\alpha}\epsilon_{\beta}e^{2(\pm n+\bar{\alpha}-\bar{\beta})\eta}\sinh\left(\frac{u}{2}\right)
-\delta_{\alpha\beta'}\cosh\left(\frac{u}{2}-2n\eta\right)\right] & \alpha \lessgtr \beta
\end{cases}
\end{equation}
\noindent
where
\begin{equation}
\epsilon_{\alpha}=\begin{cases}
1 & 1\leq\alpha\leq n\\
-1 & n+1\leq\alpha\leq 2n
\end{cases} \,,
\end{equation}

\begin{equation}
\bar{\alpha}=\begin{cases}
\alpha-\frac{1}{2} & 1\leq\alpha\leq n\\
\alpha+\frac{1}{2} & n+1\leq \alpha\leq 2n
\end{cases} \,,
\label{alphabarA2odd}
\end{equation}

\be
\alpha'=2n+1-\alpha \,,
\ee

\noindent
and $e_{\alpha\beta}$ are the $(2n) \times (2n)$ elementary matrices, 
with
\be
\alpha,\beta=1,...,2n\,.
\label{alphabeta}
\ee

\subsection{$D_{n+1}^{(2)}$}\label{sec:RmatD2}

We use the $D_{n+1}^{(2)}$ R-matrix given by Jimbo 
\cite{Jimbo:1985ua}, except we use the
variables $u$ and $\eta$ instead of $x$ and $k$, respectively, which
are related as follows:
\be
x = e^u \,, \qquad \qquad k = e^{2 \eta} \,.
\ee
We also multiply the Jimbo R-matrix by an overall factor $e^{-2u}\, 
e^{-2(n+1)\eta}$ in order to have nice crossing and 
unitarity properties. (See also \cite{Bazhanov:1984gu, 
Bazhanov:1986mu}.) Hence, this R-matrix is given by
\be
R(u)=e^{-2u}\,e^{-2(n+1)\eta}R_J(u)
\label{RD2}
\ee
with
\begin{align}
R_J(u)= & \left(e^{2u}-e^{4\eta}\right)\left(e^{2u}-e^{4n\eta}\right)\sum_{\alpha\neq n+1,n+2}e_{\alpha\alpha}\otimes e_{\alpha\alpha}
+e^{2\eta}\left(e^{2u}-1\right)\left(e^{2u}-e^{4n\eta}\right)
\sum_{\substack{\alpha\neq\beta,\beta'\\\alpha\,\textrm{or}\,\beta\neq n+1,n+2}}
\nonumber\\
& \cdot e_{\alpha\alpha}\otimes e_{\beta\beta}-
\left(e^{4\eta}-1\right)
\left(e^{2u}-e^{4n\eta}\right)
\left(\sum_{\substack{\alpha<\beta,\alpha\neq\beta' \\
\alpha,\beta\neq n+1,n+2}}+e^{2u}\sum_{\substack{\alpha>\beta,\alpha\neq \beta'\\
\alpha,\beta\neq n+1,n+2}}\right)
e_{\alpha\beta}\otimes e_{\beta\alpha}
\nonumber\\
& -\frac{1}{2}
\left(e^{4\eta}-1\right)\left(e^{2u}-e^{4n\eta}\right)
\Bigg(
\left(e^{u}+1\right)
\left(
\sum_{\alpha<n+1,\beta=n+1,n+2}+e^{u}\sum_{\alpha>n+2,\beta=n+1,n+2}
\right)
\non\\
&
\cdot\left(
e_{\alpha\beta}\otimes e_{\beta\alpha}+e_{\beta'\alpha'}\otimes e_{\alpha'\beta'}
\right)
+
\left(e^{u}-1\right)
\left(
-\sum_{\alpha<n+1,\beta=n+1,n+2}+e^{u}\sum_{\alpha>n+2,\beta=n+1,n+2}
\right)
\non\\
&
\cdot
\left(
e_{\alpha\beta}\otimes e_{\beta'\alpha}+e_{\beta'\alpha'}\otimes e_{\alpha'\beta}
\right)\Bigg)+
\sum_{\alpha,\beta\neq n+1,n+2}a_{\alpha\beta}(u)e_{\alpha\beta}\otimes e_{\alpha'\beta'}+
\frac{1}{2}\sum_{\alpha\neq n+1,n+2,\beta=n+1,n+2}
\non\\
&
\cdot
\left(
b_\alpha^{+}(u)
\left(
e_{\alpha\beta}\otimes e_{\alpha'\beta'}+e_{\beta'\alpha'}\otimes e_{\beta\alpha}
\right)
+
b_\alpha^{-}(u)
\left(
e_{\alpha\beta}\otimes e_{\alpha'\beta}+e_{\beta\alpha'}\otimes e_{\beta\alpha}
\right)
\right)
\non\\
&
+\sum_{\alpha=n+1,n+2}
\left(c^{+}(u)e_{\alpha\alpha}\otimes e_{\alpha'\alpha'}+
c^{-}(u)e_{\alpha\alpha}\otimes e_{\alpha\alpha}\right.
\non\\
&
\left. +\,
d^{+}(u)e_{\alpha\alpha'}\otimes e_{\alpha'\alpha}+
d^{-}(u)e_{\alpha\alpha'}\otimes e_{\alpha\alpha'}
\right)\,,
\label{RmatD2}
\end{align}
where for $\alpha,\beta\neq n+1,n+2$
\begin{equation}
a_{\alpha\beta}(u)=\begin{cases}
(e^{4\eta}e^{2u}-e^{4n\eta})(e^{2u}-1)& \alpha=\beta\\
(e^{4\eta}-1)(e^{4n\eta}e^{2\eta(\bar\alpha-\bar\beta)}(e^{2u}-1)-\delta_{\alpha\beta'}(e^{2u}-e^{4n\eta}))& \alpha<\beta\\
(e^{4\eta}-1)e^{2u}(e^{2\eta(\bar\alpha-\bar\beta)}(e^{2u}-1)-\delta_{\alpha\beta'}(e^{2u}-e^{4n\eta}))& \alpha>\beta
\end{cases}\,,
\end{equation}
\begin{equation}
b_{\alpha}^{\pm}(u)=\begin{cases}
\pm e^{2\eta(\alpha-1/2)}(e^{4\eta}-1)(e^{2u}-1)(e^u\pm e^{2n\eta})& \alpha<n+1\\
e^{2\eta(\alpha-n-5/2)}(e^{4\eta}-1)(e^{2u}-1)e^u(e^u\pm e^{2n\eta})& \alpha>n+2
\end{cases}\,,
\end{equation}
\be
c^{\pm}(u)=\pm\frac{1}{2}(e^{4\eta}-1)(e^{2n\eta}+1)e^u(e^u\mp 1)(e^u\pm e^{2n\eta})+e^{2\eta}(e^{2u}-1)(e^{2u}-e^{4n\eta})\,,
\ee
\be
d^{\pm}(u)=\pm\frac{1}{2}(e^{4\eta}-1)(e^{2n\eta}-1)e^u(e^u\pm 1)(e^u\pm e^{2n\eta})\,,
\ee
and
\begin{equation}
\bar{\alpha}=\begin{cases}
\alpha+1 & 1\le\alpha<n+1\\
n+\frac{3}{2} & \alpha=n+1\\
n+\frac{3}{2} & \alpha=n+2\\
\alpha-1 & n+2<\alpha\le 2n+2
\end{cases} \,,
\label{alphabarD2}
\end{equation}
\be
\alpha'=2n+3-\alpha\,.
\ee
The elementary matrices $e_{\alpha\beta}$ have dimension $(2n+2) \times (2n+2)$ with
\be
\alpha,\beta=1,\dots,2n+2\,.
\ee

\section{$D_{n+1}^{(2)}$ chains with other boundary conditions}\label{sec:other}

We briefly consider here the analytical Bethe ansatz for 
integrable open spin chains constructed using the 
$D_{n+1}^{(2)}$ R-matrix and K-matrices other than 
(\ref{KD2})-(\ref{QGKMII}), which do not have quantum group symmetry. 
We first consider the diagonal K-matrix 
(\ref{KD2diag})-(\ref{KD2diag2}), which does not have any free 
parameters,  in Sec. \ref{sec:diag}. We then 
consider in Sec. \ref{sec:blockdiag} the block-diagonal K-matrix (\ref{KD2blockparamI1})-(\ref{KD2blockparamI2})
that has a free parameter $\xi_{-}$, which reduces to case I (\ref{QGKMI}) 
when $\xi_{-}=0$. For each case with $n=1$, we propose an expression for the 
eigenvalues of the transfer matrix and the corresponding Bethe 
equations.

\subsection{Diagonal K-matrix}\label{sec:diag}

We consider here the $D_{n+1}^{(2)}$ transfer matrix (\ref{transfer}) constructed with 
the diagonal $K^{-}$-matrix (\ref{KD2diag})-(\ref{KD2diag2}), 
and with the $K^{+}$-matrix given by the automorphism (\ref{isomorphism}).
We assume that all of the corresponding eigenvalues are again given by (\ref{Lambda})
\begin{align}
\Lambda^{(m_{1}, \ldots, m_{n})}(u) &= a(u)\, A(u)\, [4 \sinh(u-2\eta) 
\sinh(u-2 n \eta) ]^{2N} \non\\
&  + \left\{\sum_{l=1}^{n} \left[ b_{l}(u)\, B_{l}(u) +
\tilde b_{l}(u)\, \tilde B_{l}(u) \right] \right\} [4 \sinh(u) 
\sinh(u-2 n \eta) ]^{2N}\non\\
&  + \tilde a(u)\, \tilde A(u)\, [4 \sinh(u) 
\sinh(u-2 (n-1) \eta) ]^{2N} \,,
\label{Lambdadiag}
\end{align}
where $A(u)$ and $B_{l}(u)$ are again given by 
(\ref{ABfuncs})-(\ref{QD2}), but the functions $a(u)$ and $b_{l}(u)$ 
are still to be determined. Arguments similar to those in Sec. 
\ref{sec:afunc} can be used to show that $a(u)$ is now given by
\be
a(u) = \frac{4 e^{2 n \eta} \cosh(u)\cosh(u-(n-1)\eta) 
\sinh(u-2n \eta) \sinh(u-(n+1)\eta)}{\sinh(2(u-\eta)) \sinh(2(u-n 
\eta))} \,,
\label{afuncdiag}
\ee
cf. (\ref{afunc}).

Let us now restrict to the simplest case $n=1$. By explicitly computing the 
eigenvalue of the transfer matrix corresponding to the reference 
state for some small value of $N$, and then comparing with $\Lambda^{(0)}(u)$ (\ref{Lambdadiag}), we learn that
$b_{1}(u)$ must satisfy the functional relation
\be
b_{1}(u) + \tilde b_{1}(u) = \frac{2 e^{2\eta} \sinh(2u) 
\sinh(2(u-2\eta))}{\sinh^{2}(2(u-\eta))} \,,
\label{b1frdiag}
\ee
where $\tilde b_{1}(u) = b_{1}(-u+2\eta)$. The ``minimal'' solution is
\be
b_{1}(u) = \tilde b_{1}(u) = \frac{e^{2\eta} \sinh(2u) 
\sinh(2(u-2\eta))}{\sinh^{2}(2(u-\eta))} \,.
\ee 
The corresponding Bethe equations are given by\footnote{The Bethe 
equations for this case have been obtained 
by Martins and Guan using coordinate Bethe 
ansatz; their result (see Eq. (48) in \cite{Martins:2000xie}) is missing the restriction ($j\ne k$) on the product, but
is otherwise equivalent to (\ref{BAEn1diag}).}
\be
e_{1}^{2N}(u_{k}^{(1)})\, e_{-1}(u_{k}^{(1)}+\frac{i\pi}{2}) = \prod_{j=1, j\ne k}^{m_{1}} 
f_{2}(u_{k}^{(1)}-u_{j}^{(1)})\, f_{2}(u_{k}^{(1)}+u_{j}^{(1)}) 
\,,\qquad k = 1, \ldots, m_{1}\,,
\label{BAEn1diag}
\ee
which have an extra factor on the LHS compared with (\ref{BAEn1}).  The
completeness of this analytical Bethe ansatz result for $N=2, 3$ is verified for a
generic value of the bulk parameter (namely, 
$\eta=-i/10$)
in Tables \ref{table:D22diag2}, \ref{table:D22diag3}, respectively, which give the solutions of the Bethe equations 
(\ref{BAEn1diag}) corresponding to each of the distinct 
transfer-matrix eigenvalues, as well as the corresponding degeneracies.
We observe that the number 
of Bethe roots $m_{1}$ lies in the interval $[0, N]$.
Evidently, the degeneracies are smaller than for the quantum-group-invariant
case, cf.  Tables \ref{table:B1N2}, \ref{table:B1N3}.

\subsection{Parameter-dependent block-diagonal K-matrix}\label{sec:blockdiag}

We now consider the block-diagonal solution 
(\ref{KD2blockparamI1})-(\ref{KD2blockparamI2}) of the $D_{n+1}^{(2)}$
BYBE (\ref{BYBEm}).
When $\xi_{-}=0$, this 
K-matrix reduces to case I  (\ref{KD2}), (\ref{QGKMI}). Moreover, we 
take $K^{+}(u)$ to be given by the isomorphism (\ref{isomorphism}), 
but with an independent boundary parameter $\xi_{+}$, i.e.
\be
K^{+}(u) = K^{-}(-u-\rho)\Big\vert_{\xi_{-} \rightarrow 
\xi_{+}}\, M\,.
\ee
It is convenient to reexpress $\xi_{\mp}$ in terms of new parameters 
$\mu_{\mp}$ as follows
\be
\xi_{-} = e^{\mu_{-} - n \eta}\,, \qquad \xi_{+} = e^{\mu_{+} + n 
\eta}\,,
\ee
which implies that the quantum-group invariant case is obtained in the limit $\mu_{\mp} \rightarrow -\infty$.

We again assume that all of the eigenvalues of the transfer matrix (\ref{transfer}) constructed with 
these K-matrices are given by
\begin{align}
\Lambda^{(m_{1}, \ldots, m_{n})}(u) &= a(u)\, A(u)\, [4 \sinh(u-2\eta) 
\sinh(u-2 n \eta) ]^{2N} \non\\
&  + \left\{\sum_{l=1}^{n} \left[ b_{l}(u)\, B_{l}(u) +
\tilde b_{l}(u)\, \tilde B_{l}(u) \right] \right\} [4 \sinh(u) 
\sinh(u-2 n \eta) ]^{2N}\non\\
&  + \tilde a(u)\, \tilde A(u)\, [4 \sinh(u) 
\sinh(u-2 (n-1) \eta) ]^{2N} \,,
\label{Lambdalockbdiag}
\end{align}
where $A(u)$ and $B_{l}(u)$ are again given by 
(\ref{ABfuncs})-(\ref{QD2}), but the functions $a(u)$ and $b_{l}(u)$ 
are still to be determined. Proceeding as in Sec. \ref{sec:afunc},
we find that $a(u)$ is now given by
\begin{align}
a(u) &= \frac{4 e^{6 n \eta}\cosh(u-n \eta) \cosh(u-(n-1)\eta) 
\sinh(2(u-2n \eta)) \sinh(u-(n+1)\eta)}{\sinh(2(u-\eta)) \sinh(u-n 
\eta)} \non\\
& \quad \times \left[-4 e^{2 n \eta+ \mu_{-} + \mu_{+}} \sinh(u+\mu_{-}) 
\sinh(u - \mu_{+} - 2 n\eta)\right] \,,
\label{afuncblockdiag}
\end{align}
which has an extra factor compared with (\ref{afunc}).

We now again restrict to the simplest case $n=1$. By explicitly computing the 
eigenvalue of the transfer matrix corresponding to the reference 
state, we find that $b_{1}(u)$ must satisfy the functional relation
\be
b_{1}(u) + \tilde b_{1}(u) = \frac{2\sinh(u)}{\sinh(u-2\eta)} 
\left[\frac{\sinh^{2}(u-\eta) - \sinh(\mu_{+} + \eta)\sinh(\mu_{-} + \eta)}{\sinh(u+\mu_{-}) 
\sinh(u - \mu_{+} - 2\eta)}\right] a(u)
\label{bsum}
\,,
\ee
where $\tilde b_{1}(u) = b_{1}(-u+2\eta)$. We proceed to solve for 
$b_{1}(u)$ by setting
\be
b_{1}(u) = \frac{\sinh(u)}{\sinh(u-2\eta)} a(u) G(u)\,,
\label{b1ansatz}
\ee
where $G(u)$ is still to be determined.
Using the fact that $\tilde a(u) = a(-u+2\eta)$ satisfies
\be
\frac{\tilde a(u)}{a(u)} = \frac{\sinh^{2}(u)}{\sinh^{2}(u-2\eta)} 
\frac{\sinh(u-\mu_{-} - 2\eta) \sinh(u + \mu_{+})}{\sinh(u+\mu_{-}) 
\sinh(u - \mu_{+} - 2\eta)} \,,
\ee
it follows from (\ref{bsum}) that $G(u)$ must satisfy the functional 
relation
\begin{align}
G(u) \sinh(u+\mu_{-})\sinh(u-\mu_{+} - 2\eta) + \tilde G(u) 
\sinh(u+\mu_{+})\sinh(u-\mu_{-} - 2\eta) \non\\
= 2 \left[\sinh^{2}(u-\eta) 
- \sinh(\mu_{+} + \eta)\sinh(\mu_{-} + \eta) \right] \,,
\label{Gfuncrltn}
\end{align}
where $\tilde G(u) = G(-u+2\eta)$. Assuming that $G(u)$ has $2i\pi$ 
periodicity and the form
\be
G(u) = \frac{G^{(1)}(u)}{G^{(2)}(u)} \quad \mbox{ where } \quad G^{(j)}(u) = 
\sum_{k=-1}^{1} g^{(j)}_{k} e^{k u}\,,\quad j = 1, 2, 
\ee
we find the minimal solution\footnote{We remark that 
(\ref{Gfunc}) satisfies $G(-u+2\eta)=1/G(u)$. With this ansatz, the functional relation 
(\ref{Gfuncrltn}) becomes a quadratic equation for $G(u)$, which has 
(\ref{Gfunc}) as one of its two solutions.}
\be
G(u) = \frac{\cosh(\frac{1}{2}(u+\mu_{+})) 
\cosh(\frac{1}{2}(u-\mu_{-}-2\eta))}{\cosh(\frac{1}{2}(u+\mu_{-})) 
\cosh(\frac{1}{2}(u-\mu_{+}-2\eta))} \,.
\label{Gfunc}
\ee
In short, we take $b_{1}(u)$ to be given by (\ref{b1ansatz}) and 
(\ref{Gfunc}).

The Bethe equations, which we obtain from the expression for 
$\Lambda^{(m_{1})}(u)$ (\ref{Lambdalockbdiag})
by demanding the cancellation of the residues from the poles in 
$A(u)$ and $B_{1}(u)$ at $u=u_{k}^{(1)} + \eta$, are given by
\begin{align}
& a(u_{k}^{(1)}+ \eta) \Bigg[e_{1}^{2N}(u_{k}^{(1)})\, 
f_{1+\mu_{+}/\eta}(u_{k}^{(1)}+i\pi) f_{-1-\mu_{-}/\eta}(u_{k}^{(1)}+i\pi) \non\\
& \qquad\qquad\qquad - \prod_{j=1, j\ne k}^{m_{1}} 
f_{2}(u_{k}^{(1)}-u_{j}^{(1)})\, f_{2}(u_{k}^{(1)}+u_{j}^{(1)}) 
\Bigg] = 0 \,, \qquad k = 1, \ldots, m_{1}\,,
\label{preBAEn1blockdiag}
\end{align} 
where the notation is defined in (\ref{defsef}). Assuming that the 
prefactor $a(u_{k}^{(1)}+ \eta)$ is nonzero, the above Bethe 
equations reduce to a more conventional form
\begin{align}
& e_{1}^{2N}(u_{k}^{(1)})\, 
f_{1+\mu_{+}/\eta}(u_{k}^{(1)}+i\pi) f_{-1-\mu_{-}/\eta}(u_{k}^{(1)}+i\pi) = \prod_{j=1, j\ne k}^{m_{1}} 
f_{2}(u_{k}^{(1)}-u_{j}^{(1)})\, f_{2}(u_{k}^{(1)}+u_{j}^{(1)}) \,, \non \\
&\qquad\qquad\qquad\qquad\qquad k = 1, \ldots, m_{1}\,.
\label{BAEn1blockdiag}
\end{align} 
The completeness of this analytical Bethe ansatz result for $N=1, 2$ is verified for 
generic values of the bulk and boundary parameters (namely, 
$\eta=-i/10$, $\mu_{-}=1/5$, $\mu_{+}=1/7$), 
in Tables \ref{table:D22blockdiag1}, \ref{table:D22blockdiag2}, 
respectively, which give the solutions of the Bethe equations 
(\ref{BAEn1blockdiag}) corresponding to each of the transfer-matrix eigenvalues.
Note that the number 
of Bethe roots $m_{1}$ now lies in the interval $[0, 2N]$.  There are 
now no degeneracies.

\subsection{Special manifold}

We remark that on the ``special manifold'' $\mu_{-} = 
\mu_{+} \equiv \mu$, i.e. $\xi_{+} = e^{2\eta} \xi_{-}$, 
we have that $G(u)=1$ and $b_{1}(u) = \tilde b_{1}(u) = 
\frac{\sinh(u)}{\sinh(u-2\eta)} a(u)$; hence, the expression for 
$\Lambda^{(m_{1})}(u)$ is proportional to the one for the 
quantum-group invariant case $\mu_{\mp} \rightarrow -\infty$. 
However, contrary to the claim in \cite{Martins:2000xie}, the Bethe 
equations do {\em not} reduce to those of the 
quantum-group invariant case (\ref{BAEn1}). Indeed, 
the assumption used to 
pass from (\ref{preBAEn1blockdiag}) to (\ref{BAEn1blockdiag})
(namely, the nonvanishing of $a(u_{k}^{(1)}+ \eta)$) is no 
longer valid for this particular case. Hence, the Bethe equations on the 
``special manifold'' are given by
\be
a(u_{k}^{(1)}+ \eta) \Bigg[e_{1}^{2N}(u_{k}^{(1)})\, 
 - \prod_{j=1, j\ne k}^{m_{1}} 
f_{2}(u_{k}^{(1)}-u_{j}^{(1)})\, f_{2}(u_{k}^{(1)}+u_{j}^{(1)}) 
\Bigg] = 0 \,, \qquad k = 1, \ldots, m_{1}\,.
\label{BAEn1special}
\ee 
In other words, there are two Bethe equations on the 
``special manifold'' (either of which must be satisfied): the Bethe equations for the quantum-group 
invariant case (\ref{BAEn1}), and $a(u_{k}^{(1)}+ \eta) = 0$. The latter 
has the solution 
\be
u_{k}^{(1)} = \mu + \eta + i \pi l\,, \qquad  l \in \mathbb{Z}\,. 
\label{specialBetheroots}
\ee
This solution (which 
evidently depends on the value of the boundary parameter $\mu$) must be included in order to 
obtain the complete spectrum.   In 
hindsight, since the spectrum on the ``special manifold'' is not the 
same as for the quantum-group invariant case, it should come as no 
surprise that the Bethe equations for these two cases are not the same.

Completeness on the ``special manifold'' for $N=1, 2$ is verified  
(for $\eta=-i/10$, $\mu_{-}=\mu_{+}=1/5$), 
in Tables \ref{table:D22special1}, \ref{table:D22special2}, 
respectively. Solutions that contain the special Bethe root (\ref{specialBetheroots}) are 
denoted by a dagger (${}^{\dagger}$). We observe that there are many 
such solutions. Moreover, solutions that do {\em not} contain this 
special root must also be solutions of (\ref{BAEn1}); and, indeed, 
the solutions in Table \ref{table:D22special2} without 
a dagger also appear in the corresponding 
Table  \ref{table:B1N2} for the quantum-group invariant chain. We 
are not aware of other examples of ``hybrid'' Bethe ansatz systems 
like (\ref{BAEn1special}), which achieve completeness in such a 
curious fashion.

\newpage
\clearpage

\begin{table}
\centering
\begin{tabular}{|c|c|c|c|c|}
\hline
$m_{1}$ & $a_{1}$ & deg & mult & $\{ u^{(1)}_{k} \}$\\   
\hline
0 & 2  & 3 & 1 & - \\
\hline
1 & 0  & 1 & 1 & 0.205557 \\
\hline
\end{tabular}
\caption{\small $A_{1}^{(2)}$ - case I with
$U_{q}(C_{1})$ symmetry, $N=2$}\label{table:C1N2}
\end{table}

\begin{table}
\centering
\begin{tabular}{|c|c|c|c|c|}
\hline
$m_{1}$ & $a_{1}$ & deg & mult & $\{ u^{(1)}_{k} \}$\\   
\hline
0 & 3  & 4 & 1 & - \\
\hline
1 & 1  & 2 & 2 &  0.117573 \\
  &    &   &   &  0.366703 \\
\hline
\end{tabular}
\caption{\small $A_{1}^{(2)}$ - case I with
$U_{q}(C_{1})$ symmetry, $N=3$}\label{table:C1N3}
\end{table}

\begin{table}
\centering
\begin{tabular}{|c|c||c|c||c|c|c|c|}
\hline
$m_{1}$ & $m_{2}$ & $a_{1}$ & $a_{2}$ & deg & mult & $\{ u^{(1)}_{k} 
\}$ & $\{ u^{(2)}_{k} \}$\\   
\hline
0 & 0 & 2 & 0  & 10 & 1 & - & - \\
\hline
1 & 0 & 0 & 1  & 5 & 1 & 0.201347  & - \\
\hline
2 & 1 & 0 & 0  & 1 & 1 & 0.210433, $1.57268 i$  &  $0.760991 i$\\
\hline
\end{tabular}
\caption{\small $A_{3}^{(2)}$ - case I with
$U_{q}(C_{2})$ symmetry, $N=2$}\label{table:C2N2}
\end{table}

\begin{table}
\centering
\begin{tabular}{|c|c||c|c||c|c|c|c|}
\hline
$m_{1}$ & $m_{2}$ & $a_{1}$ & $a_{2}$ & deg & mult & $\{ u^{(1)}_{k} 
\}$ & $\{ u^{(2)}_{k} \}$\\   
\hline
0 & 0 & 3 & 0  & 20 & 1 & - & - \\
\hline
1 & 0 & 1 & 1  & 16 & 2 & 0.115986  & - \\
  &   &   &    &    &   & 0.351133  & - \\
\hline
2 & 1 & 1 & 0  & 4 & 3 & 0.117014, 0.361311  &  0.345671\\
  &   &   &    &   &   & 0.118818, $1.76308  i$ & $2.38373 i$\\
  &   &   &    &   &   & 0.380307, $1.80836 i$ & $0.67199 i$\\
  \hline
\end{tabular}
\caption{\small $A_{3}^{(2)}$ - case I with
$U_{q}(C_{2})$ symmetry, $N=3$}\label{table:C2N3}
\end{table}

\begin{table}
\small 
\centering
\begin{tabular}{|c|c|c||c|c|c||c|c|c|c|c|}
\hline
$m_{1}$ & $m_{2}$ & $m_{3}$ & $a_{1}$ & $a_{2}$  & $a_{3}$ & deg & mult & $\{ u^{(1)}_{k} 
\}$ & $\{ u^{(2)}_{k} \}$ & $\{ u^{(3)}_{k} \}$\\   
\hline
0 & 0 & 0 & 2 & 0 & 0  & 21 & 1 & - & - & - \\
\hline
1 & 0 & 0 & 0 & 1 & 0  & 14 & 1 & 0.201347  & - & - \\
\hline
2 & 2 & 1 & 0 & 0 & 0  & 1 & 1 & 0.216671, $1.20705 i$
& $0.584376 i\,, 1.70153 i$  &  $0.944039 i$\\
\hline
\end{tabular}
\caption{\small $A_{5}^{(2)}$ - case I with
$U_{q}(C_{3})$ symmetry, $N=2$}\label{table:C3N2}
\end{table}

\begin{table}
\small 
\centering
\begin{tabular}{|c|c|c||c|c|c||c|c|c|c|c|}
\hline
$m_{1}$ & $m_{2}$ & $m_{3}$ & $a_{1}$ & $a_{2}$  & $a_{3}$ & deg & mult & $\{ u^{(1)}_{k} 
\}$ & $\{ u^{(2)}_{k} \}$ & $\{ u^{(3)}_{k} \}$\\   
\hline
0 & 0 & 0 & 3 & 0 & 0  & 56 & 1 & - & - & - \\
\hline
1 & 0 & 0 & 1 & 1 & 0  & 64 & 2 & 0.115986  & - & - \\
  &   &   &   &   &    &    &   & 0.351133  & - & - \\
\hline
2 & 1 & 0 & 0 & 0 & 1  & 14 & 1 & 0.115986, 0.351133
& 0.331791  &  - \\
\hline
2 & 2 & 1 & 1 & 0 & 0  & 6 & 3 & 0.118139, 0.373599
& 0.362631, $1.57735 i$  &  $0.710525 i$\\
  &   &   &   &   &    &    &   & 0.399729, $1.4295 i$ & $0.480282  
  i\,, 1.83749  i$ & $0.863899 i$ \\
&   &   &   &   &    &    &   & 0.1203946, $1.3871 i$ & $0.613358  
  i\,, 1.78304  i$ & $0.943811 i$ \\  
\hline
\end{tabular}
\caption{\small $A_{5}^{(2)}$ - case I with
$U_{q}(C_{3})$ symmetry, $N=3$}\label{table:C3N3}
\end{table}

\begin{table}
\centering
\begin{tabular}{|c|c|c|c|c|}
\hline
$m_{1}$ & deg & $\{ u^{(1)}_{k} \}$\\   
\hline
0 & 2 &  - \\
\hline
1 & 1 & 0.197385 \\
  & 1 & $0.867208 + 1.5708 i$ \\
\hline
\end{tabular}
\caption{\small $A_{1}^{(2)}$ - case II, $N=2$}\label{table:D1N2}
\end{table}

\begin{table}
\centering
\begin{tabular}{|c|c|c|c|c|}
\hline
$m_{1}$ & deg & $\{ u^{(1)}_{k} \}$\\   
\hline
0 & 2 &  - \\
\hline
1 & 2 & 0.115455 \\
  & 2 & 0.346805 \\
  & 2 & $0.643153 + 1.5708 i$ \\
\hline
\end{tabular}
\caption{\small $A_{1}^{(2)}$ - case II, $N=3$}\label{table:D1N3}
\end{table}

\begin{table}
\centering
\begin{tabular}{|c|c||c|c||c|c|c|c|}
\hline
$m_{1}$ & $m_{2}$ & $a_{1}$ & $a_{2}$ & deg & mult & $\{ u^{(1)}_{k} 
\}$ & $\{ u^{(2)}_{k} \}$\\   
\hline
0 & 0 & 2 & 2  & 9 & 1 & - & - \\
\hline
1 & 0 & 0 & 2  & 6 (*) &   & 0.201347  & - \\
\hline
2 & 1 & 0 & 0  & 1 & 1 & $0.504878 \pm 1.10246 i$  &  $0.623371 + 1.5708  i$\\
\hline
\end{tabular}
\caption{\small $A_{3}^{(2)}$ - case II with
$U_{q}(D_{2})$ symmetry, $N=2$}\label{table:D2N2}
\end{table}

\begin{table}
\centering
\begin{tabular}{|c|c||c|c||c|c|c|c|}
\hline
$m_{1}$ & $m_{2}$ & $a_{1}$ & $a_{2}$ & deg & mult & $\{ u^{(1)}_{k} 
\}$ & $\{ u^{(2)}_{k} \}$\\   
\hline
0 & 0 & 3 & 3  & 16 & 1 & - & - \\
\hline
1 & 0 & 1 & 3  & 16 (*) &   & 0.115986  & - \\
  &   &   &    & 16 (*) &   & 0.351133  & - \\
\hline
2 & 1 & 1 & 1  & 4 & 4 & 0.115986, 0.351133  &  0.324295\\
  &   &   &    &   &   & 0.114078, $1.20875 i$  & $0.536882 + 1.5708 i$ \\
  &   &   &    &   &   & 0.340353, $1.30112 i$  & $0.474472 + 1.5708 i$ \\
  &   &   &    &   &   & $0.447272 \pm 1.23578 i$  & $0.509118 + 
  1.5708 i$ \\
\hline
\end{tabular}
\caption{\small $A_{3}^{(2)}$ - case II with
$U_{q}(D_{2})$ symmetry, $N=3$}\label{table:D2N3}
\end{table}

\begin{table}
\small 
\centering
\begin{tabular}{|c|c|c||c|c|c||c|c|c|c|c|}
\hline
$m_{1}$ & $m_{2}$ & $m_{3}$ & $a_{1}$ & $a_{2}$  & $a_{3}$ & deg & mult & $\{ u^{(1)}_{k} 
\}$ & $\{ u^{(2)}_{k} \}$ & $\{ u^{(3)}_{k} \}$\\   
\hline
0 & 0 & 0 & 2 & 0 & 0  & 20 & 1 & - & - & - \\
\hline
1 & 0 & 0 & 0 & 1 & 1  & 15 & 1 & 0.201347  & - & - \\
\hline
2 & 2 & 1 & 0 & 0 & 0  & 1 & 1 & $0.371298 \pm 0.870141i$
& $0.429596 \pm 1.24327 i$  &  $0.490237 + 1.5708 i$\\
\hline
\end{tabular}
\caption{\small $A_{5}^{(2)}$ - case II with
$U_{q}(D_{3})$ symmetry, $N=2$}\label{table:D3N2}
\end{table}

\begin{table}
\small 
\centering
\begin{tabular}{|c|c|c||c|c|c||c|c|c|c|c|}
\hline
$m_{1}$ & $m_{2}$ & $m_{3}$ & $a_{1}$ & $a_{2}$  & $a_{3}$ & deg & mult & $\{ u^{(1)}_{k} 
\}$ & $\{ u^{(2)}_{k} \}$ & $\{ u^{(3)}_{k} \}$\\   
\hline
0 & 0 & 0 & 3 & 0 & 0  & 50 & 1 & - & - & - \\
\hline
1 & 0 & 0 & 1 & 1 & 1  & 64 & 2 & 0.115986  & - & - \\
  &   &   &   &   &    &    &   & 0.351133  & - & - \\
\hline
2 & 1 & 0 & 0 & 0 & 2  & 20 (*) &   & 0.115986, 0.351133
& 0.331791  & -\\
\hline
2 & 2 & 1 & 0 & 0 & 0  & 6 & 3 & 0.111983, $0.878547 i$
& $0.337611 \pm 1.1629  i$  &  $0.468715 + 1.5708 i$\\
  &   &   &   &   &    &    &   & 0.336241, $0.995464 i$  & $0.254738 
  \pm 1.16415 i$ & $0.416013 + 1.5708 i$ \\
  &   &   &   &   &    &    &   & $0.345988 \pm 1.01689 i$   & 
  $0.378084 \pm 1.3022 i$ & $0.413388 + 1.5708 i$ \\
\hline
\end{tabular}
\caption{\small $A_{5}^{(2)}$ - case II with
$U_{q}(D_{3})$ symmetry, $N=3$}\label{table:D3N3}
\end{table}

\begin{table}
\small
\centering
\begin{tabular}{|c|c|c|c|c|}
\hline
$m_{1}$ & $a_{1}$ & deg & mult & $\{ u^{(1)}_{k} \}$\\  
\hline
0 & 4 & 5 & 1  & - \\
\hline
1 & 2  &   3    & &  $1.5708 i$ \\
  &    & 6  (*) & & $0.100673 $ \\  % Here l=n=1. But since m_{1} is NOT >1, can shift by i \pi 
\hline
2  & 0 & 1 & 2  & $0.100167$\,, $0.100167 + 3.14159 i$ \\
   &   &   &    & $0.545151\pm 1.5708i$\\
\hline
\end{tabular}
\caption{\small $D_{2}^{(2)}$ - case I with
$U_{q}(B_{1})$ symmetry, $N=2$}\label{table:B1N2}
\end{table}

\begin{table}
\small
\centering
\begin{tabular}{|c|c|c|c|c|}
\hline
$m_{1}$ & $a_{1}$ & deg & mult & $\{ u^{(1)}_{k} \}$\\   
\hline
0  & 6 & 7  & 1  & - \\
\hline
1  & 4 & 5 &  & $1.5708 i$  \\
 &  & 10  (*) &  & $0.0579932$ \\  % Here l=n=1. But since m_{1} is NOT >1, can shift by i \pi                                         
 &  & 10  (*) & & $0.175567$  \\  % Here l=n=1. But since m_{1} is NOT >1, can shift by i \pi
\hline
2  & 2  & 3  &  & $0.428774 \pm 1.5708 i$ \\
&  & 3     & & $0.174378\,$, $0.174378\, +3.14159 i$  \\
&  & 3    & & $0.0578635\,$,$0.0578635 + 3.14159 i$ \\
&  & 6   (*) & & $1.06487 i\,$, $0.171045$ \\
&  & 6   (*) & & $1.04228 i\,$, $0.0574593$ \\
&  & 6   (*) & & $0.17437\,$, $0.0578644\, +3.14159 i$ \\
\hline
3  & 0  & 1  &  & $0.873189 \pm 1.5708 i$\,, $1.5708 i$  \\
 &  & 1  &  & 0.172061\,, $0.17206 + 3.14159 i$\,, $1.5708 i$ \\
 &  & 1   & & $0.0576038\,$,$0.0576038 + 3.14159 i\,$,$1.5708 i$ \\
 &  & 2 (*) &  & 0.172083\,,$1.58398 i$\,,  $0.0576011 + 3.14159 i$ \\
\hline
\end{tabular}
\caption{\small $D_{2}^{(2)}$ - case I with
$U_{q}(B_{1})$ symmetry, $N=3$}\label{table:B1N3}
\end{table}

\begin{table}
\small
\centering
\begin{tabular}{|c|c||c|c||c|c|c|c|}
\hline
$m_{1}$ & $m_{2}$ & $a_{1}$ & $a_{2}$ & deg & mult & $\{ u^{(1)}_{k} \}$ & $\{ u^{(2)}_{k} \}$\\   
\hline
0 & 0 & 2 & 0  & 14 & 1 & - & - \\
\hline
1 & 0 & 0 & 2  & 10 & 1 & 0.100673  & - \\
\hline
1 & 1 & 1 & 0  & 5 & 2 & 0.10171  &  $1.5708 i$\\
 &  &  &   &  &  & $0.954127 i$  &  $1.5708 i$\\
\hline
2 & 2 & 0 & 0  & 1 & 2 & 0.0838294, 0.234086  &  0.202721, $0.202721 + 
3.14159 i$\\
 &  &  &   &  &  & $0.319555 \pm 0.866663 i$    & $0.425346 \pm 
 1.5708 i$\\
\hline
\end{tabular}
\caption{\small $D_{3}^{(2)}$ - case I with
$U_{q}(B_{2})$ symmetry, $N=2$}\label{table:B2N2}
\end{table}

\begin{table}
\begin{adjustwidth}{-1cm}{}
\small
\centering
\begin{tabular}{|c|c||c|c||c|c|c|c|}
\hline
$m_{1}$ & $m_{2}$ & $a_{1}$ & $a_{2}$ & deg & mult & $\{ u^{(1)}_{k} 
\}$ & $\{ u^{(2)}_{k} \}$\\   
\hline
0 & 0 & 3 & 0  & 30 & 1 & - & - \\
\hline
1 & 0 & 1 & 2  & 35 & 2 & 0.0579932  & - \\
 &  &  &   &  &  & 0.175567  & - \\
\hline
1 & 1 & 2 & 0  & 14 & 3 & 0.0582547  & $1.5708 i$ \\
 &  &  &   &  &  & 0.178031  & $1.5708 i$ \\
 &  &  &   &  &  & $1.04525 i$  & $1.5708 i$ \\
\hline
2 & 1 & 0 & 2  & 10 &   & 0.168984\,, $0.804061 i$  & $1.5708 i$ \\
 &  &  &   & 10 &   & 0.0571605\,, $0.778499 i$  & $1.5708 i$ \\
 &  &  &   & 20 (*) &   & 0.0579932\,, 0.175567  & 0.165895 \\
\hline
2 & 2 & 1 & 0  & 5 & 6 & 0.0570691\,, $0.693318  i$  & $0.478036 \pm 
1.5708 i$ \\
 &  &  &   &  &  & $0.277208  \pm 1.01515 i$ & $0.360095  \pm 1.5708 i$ \\
 &  &  &   &  &  &  0.168801\,, $0.726058 i$ & $0.473141 \pm 1.5708 i$ \\
 &  &  &   &  &  &  0.148227\,, 0.321648 & 0.270361\,, $0.270361 + 3.14159 i$ \\
 &  &  &   &  &  &  0.0544035\,, 0.306275 & 0.242382\,, $0.242382  + 3.14159 i$ \\
 &  &  &   &  &  &  0.0500765\,, 0.120433 & 0.136275\,, $0.136275 + 3.14159 i$\\
\hline
3 & 3 & 0 & 0  & 1 & 4 & $0.551099 \pm 0.913068 i$\,, $0.946389  i$  
& $0.696506 \pm 1.5708 i$\,, $1.5708 i$ \\
&  &  &   &  &  & 0.144293\,, 0.308622\,, $0.811225 i$ & 0.264603\,, $1.5708 i$\,, $0.264603 + 3.14159 i$\\
&  &  &   &  &  & 0.0537237\,, 0.294364\,, $0.798067 i$ & 0.238335\,, 
$0.238335 + 3.14159 i$\,, $1.5708 i$\\
&  &  &   &  &  & 0.049605\,, 0.11872\,, $0.762414  i$ & 0.136965\,, $0.136965 + 3.14159 i$\,, $1.5708 i$\\ 
\hline
\end{tabular}
\caption{\small $D_{3}^{(2)}$ - case I with
$U_{q}(B_{2})$ symmetry, $N=3$}\label{table:B2N3}
\end{adjustwidth}
\end{table}

\begin{table}
\centering
\begin{tabular}{|c|c|c|}
\hline
$m_{1}$ & deg & $\{ u^{(1)}_{k} \}$\\   
\hline
0 & 2  &  - \\
1 & 4  & 0.545151 + $1.5708 i$ \\
1 & 4  & 0.100167 \\
2 & 2  & 0.397493 + $1.5708i$, 1.6011 + $1.5708 i$  \\
2 & 2  & $0.126817 \pm 0.0788484 i $  \\
2 & 1  & $0.877837 \pm 1.5708  i $  \\
2 & 1  & 0.0996683, 0.0996683 + $3.14159 i$ \\
\hline
\end{tabular}
\caption{\small $D_{2}^{(2)}$ with
diagonal K-matrices, $N=2$}\label{table:D22diag2}
\end{table}

\begin{table}
\centering
\begin{tabular}{|c|c|c|}
\hline
$m_{1}$ & deg & $\{ u^{(1)}_{k} \}$\\   
\hline
0 & 2  &  - \\
1 & 4  & 0.428774 + $1.5708 i$ \\
1 & 4  & 0.174378 \\
1 & 4  & 0.0578635 \\
2 & 4  & 1.01089 - $1.5708i$, 0.26304 - $1.5708 i$  \\
2 & 4  & 0.470322 + $1.27019 i $, 0.172028 - $0.00158932 i$   \\
2 & 4  & 0.470322 + $1.8714 i $, 0.172028 - $ 3.14 i$ \\
2 & 4  & 0.0576039 + $0.000187623 i$, 0.474343 - $1.2624 i$ \\
2 & 4  & 0.0576039 - $0.000187623 i$, 0.474343 + $1.2624 i $ \\
2 & 4  & 0.17321, 0.0577354 + $3.14159 i$  \\
2 & 2  & 0.173217, 0.173217 + $3.14159 i$ \\
2 & 2  & $0.654634  \pm 1.5708 i$ \\
2 & 2  & 0.0577346, 0.0577346 + $3.14159 i$ \\
3 & 2  & 0.879251  - $1.5708 i$, 1.49011  + $1.5708 i $, 0.44759  + $1.5708 i $  \\
3 & 2  & 0.862697  + $1.5708 i$, 2.15526  + $1.5708 i $, 0.216453 + $1.5708 i $  \\
3 & 2  & 0.172024 - $ 0.00228577 i$, 1.5635 + $ 0.80969 i$, 0.298597 
+ $ 1.28321 i$\\
3 & 2  & 1.5635 - 0.80969 i$, 0.172024 + $ 0.00228577 i$, 0.298597 - 
1.28321 i$\\
3 & 2  & 1.57111 - $ 0.81849  i$, 0.298184 - $1.27838 i$, 0.0576005 
+$ 0.000261013 i$\\
3 & 2  & 1.57111 + $ 0.81849 i$, 
0.298184 + $ 1.27838 i$, 0.0576005 - $ 0.000261013 i$\\
3 & 2 & 0.172036 - $0.00113145 i$, 0.172036 - $3.14046 i$, 0.863851 + 
$1.5708 i$  \\
3 & 2 & 0.172044 + $0.00113171 i$, 0.0576064 + $3.14146 i$, 0.873191 
-$ 1.57093 i$\\
3 & 2 & 0.172044 - $0.00113171 i$, 0.0576064 - $3.14146 i$, 0.873191 
+ $ 1.57093 i$ \\
3 & 2 & 0.0576073 - $0.000129149 i$, 0.0576073 - $3.14146 i$, 
0.882594 + $1.5708 i$  \\
\hline
\end{tabular}
\caption{\small $D_{2}^{(2)}$ with
diagonal K-matrices, $N=3$}\label{table:D22diag3}
\end{table}

\begin{table}
\centering
\begin{tabular}{|c|c|c|}
\hline
$m_{1}$ & deg & $\{ u^{(1)}_{k} \}$\\   
\hline
0 & 1  &  - \\
1 & 1  & 0.137015 - $1.54958 i$ \\
1 & 1  & 0.182573 + $3.06075 i$ \\
2 & 1  & 0.182446 + $3.06091 i$, 2.65675 - $1.44564 i$  \\
\hline
\end{tabular}
\caption{\small $D_{2}^{(2)}$ with parameter-dependent
block-diagonal K-matrices, $N=1$}\label{table:D22blockdiag1}
\end{table}

\begin{table}
\centering
\begin{tabular}{|c|c|c|}
\hline
$m_{1}$ & deg & $\{ u^{(1)}_{k} \}$\\   
\hline
0 & 1  &  - \\
1 & 1  & 0.100674 - $0.0000724644  i$ \\
1 & 1  & 0.0936584 - $3.13986  i$ \\
1 & 1  & 0.188618 + $3.03964 i$ \\
1 & 1  & 0.0698465 - $1.56476 i$ \\
2 & 1  & 0.100167 - $0.0000713337 i$, 0.0932544 - $3.13987  i$  \\
2 & 1  & 0.10017 - $0.0000762043 i$, 0.188722 + $3.03998 i$  \\
2 & 1  & 0.505942 - $2.60599  i$, 0.0934598 + $3.14034 i$  \\
2 & 1  & 0.454856 - $3.1392  i$,  0.186277 + $3.03322i$  \\
2 & 1  & 0.187084 + $3.0415  i$,  0.380464 - $2.47146i$  \\
2 & 1  & 0.424372 + $1.57794  i$, 0.664226 - $1.53054 i$  \\
3 & 1  & 0.452946 - $3.03685 i$,  0.184102 + $3.03559 i$, 0.0996929 - $0.0000640892 i$  \\
3 & 1  & 0.464307 - $2.57342 i$,  0.0930099 + $3.1403i$, 3.33685 - $1.36781  i$  \\
3 & 1  & 0.445549 - $3.13242 i$,  0.1859 + $3.03306i$, 3.38217 - $1.51649 i$  \\
3 & 1  & 0.187283 + $3.04163 i$,  0.349451 - $2.43968i$, 3.27327 - $1.37062 i$  \\
4 & 1  & 0.439821 - $3.0445  i$,  0.183825 + $3.03521i$, 1.60465 - 
$0.89409 i$, 1.74283 + $0.606952 i$  \\
\hline
\end{tabular}
\caption{\small $D_{2}^{(2)}$ with parameter-dependent
block-diagonal K-matrices, $N=2$}\label{table:D22blockdiag2}
\end{table}

\begin{table}
\centering
\begin{tabular}{|c|c|c|}
\hline
$m_{1}$ & deg & $\{ u^{(1)}_{k} \}$\\   
\hline
0 & 1  &  - \\
1 & 1  & $1.5708 i$ \\
1${}^{\dagger}$ & 2  & 0.2 + $3.04159  i$ \\
\hline
\end{tabular}
\caption{\small $D_{2}^{(2)}$ on the ``special manifold'', 
$N=1$}\label{table:D22special1}
\end{table}

\begin{table}
\centering
\begin{tabular}{|c|c|c|}
\hline
$m_{1}$ & deg & $\{ u^{(1)}_{k} \}$\\   
\hline
0 & 1  &  - \\
1 & 2  & 0.100673 + $3.14159 i$ \\
1 & 1  & $1.5708 i$ \\
1${}^{\dagger}$ & 2  & 0.2 - $ 0.1 i$ \\
2 & 1  & 0.100167, 0.100167 + $3.14159 i$  \\
2 & 1  &  $0.545151 \pm 1.5708 i$  \\
2${}^{\dagger}$ & 2  & 0.10017 + $3.14159 i$, 0.2 - $0.1 i$  \\
2${}^{\dagger}$ & 2  & 0.123242 + $0.0334665 i$,  0.2 - $0.1 i$  \\
2${}^{\dagger}$ & 2  & 0.614679 - $0.205195 i$,  0.2 - $0.1 i$  \\
2${}^{\dagger}$ & 2  & 0.151836 + $0.629616 i$,  0.2 - $0.1 i$  \\
\hline
\end{tabular}
\caption{\small $D_{2}^{(2)}$ on the ``special manifold'', 
$N=2$}\label{table:D22special2}
\end{table}

\newpage
\clearpage

% \bibliographystyle{utphys}
% \bibliography{refs}

\providecommand{\href}[2]{#2}\begingroup\raggedright\endgroup

\end{document}